\documentclass[review]{elsarticle}
\usepackage{amsmath,amssymb}
\usepackage{lipsum}
\usepackage{bm}
\usepackage{graphicx}
\usepackage{graphics}
\usepackage{amsmath}
\usepackage{array}
\usepackage[caption=false]{subfig}
\usepackage{xcolor}
\usepackage{subfig}
\usepackage{hyperref}
\usepackage[capitalise]{cleveref}

\usepackage{textcomp} 

\usepackage{epstopdf}
\usepackage{cases}
\usepackage{amssymb}
\usepackage{multirow}
\usepackage{siunitx}
\usepackage{cases}
\usepackage{tabularx}

\usepackage[top=1.1in, bottom=1.1in, left=1.0in, right=1.0in]{geometry}
\usepackage[utf8]{inputenc}
\usepackage{tikz}
\usepackage{comment}
\usepackage{lineno,hyperref}
\modulolinenumbers[5]
\usepackage{booktabs}

\begin{document}
\title{An implicit kinetic scheme for multiscale heat transfer problem accounting for phonon dispersion and polarization}
\author[add1]{Chuang Zhang}
\ead{zhangcmzt@hust.edu.cn}
\author[add1]{Songze Chen}
\ead{jacksongze@hust.edu.cn}
\author[add1]{Zhaoli Guo\corref{cor1}}
\ead{zlguo@hust.edu.cn}
\cortext[cor1]{Corresponding author}
  \address[add1]{State Key Laboratory of Coal Combustion, Huazhong University of Science and Technology,Wuhan, 430074, China}

\date{\today}
\begin{abstract}

An efficient implicit kinetic scheme is developed to solve the stationary phonon Boltzmann transport equation (BTE) based on the non-gray model including the phonon dispersion and polarization.
Due to the wide range of the dispersed phonon mean free paths, the phonon transport under the non-gray model is essentially multiscale, and has to be solved differently and appropriately for varied phonon frequencies and branches.
The proposed implicit kinetic scheme is composed of a microscopic iteration and a macroscopic iteration. The microscopic iteration is capable of automatically adapting with varied phonon mean free path of each phonon frequency and branch through solving the phonon BTE.
The energy transfer of all phonons is gathered together by the microscopic iteration to evaluate the heat flux.
The temperature field is predicted through a macroscopic heat transfer equation according to the heat flux, and the equilibrium state in the phonon BTE is also updated.
The combination of the phonon BTE solver and the macroscopic equation makes the present method very efficient in a wide length scale.
Three numerical tests, including the cross-plane, in-plane and nano-porous heat transfer in silicon, validate that the present scheme can handle with the phonon dispersion and polarization correctly and predict the multiscale heat transfer phenomena efficiently in a wide range.
The present method could be tens of times faster than the typical implicit DOM and keeps the same amount of the memory requirements as the Fourier solver for multiscale heat transfer problem.

\end{abstract}
\maketitle
\section{Introduction}

With the development of semiconductor materials and micro-nano technology, the multiscale heat transfer problem is intensively investigated in the past decades.
This kind of problem is generally featured by the non-Fourier heat transfer phenomenon~\cite{toberer2012advances,Minnich15advances,cahill2014nanoscale,cahill2003nanoscale} which can be described by the phonon Boltzmann transport equation (BTE)~\cite{zhangZm07HeatTransfer,ZimanJM60phonons,ChenG05Oxford,Majumdar98MET}.
Since the mean free path of phonon with different frequencies varies in a wide range, the heat transfer problem described by the phonon BTE is essentially multiscale with complex interactions among various phonons.

In order to tackle with varied phonon frequencies and branches, several models of the phonon BTE are developed for phonon transport~\cite{MurthyJY05Review,narumanchi2005comparison,Minnich15advances,narumanchi2004submicron}, such as gray model, semi-gray model and non-gray model.
In the gray model, all phonons transport with single frequency, group velocity and relaxation time.
Under this model, some primary non-Fourier heat transfer phenomena and boundary scattering effects are successfully reproduced, for example the temperature slip.
However, it cannot represent the granularity of phonon behavior, including the contribution of each phonon branch to the thermal conduction.
This leads to significant discrepancies with experimentally observed behaviors under many circumstances~\cite{ju1999phonon,cahill2014nanoscale,cahill2003nanoscale}.
Another relatively sophisticated model is the semi-gray model, or "two-fluid model"~\cite{ju1999microscale,Armstrong81twofluid}, in which all phonons are separated into two parts: the reservoir group and the propagating group.
The former is responsible for the thermal capacitance while the latter responsible for the thermal transport.
However, it is well known that real phonon dispersion relations are nonlinear in silicon~\cite{holland1963analysis,brockhouse1959lattice} and the phonon transport with a wide range of frequencies.
The simplification in the frequency space makes above two models simple and inexpensive,
but powerless to describe the highly non-equilibrium heat transfer problem in real thermal application.
Actually, a lot of studies~\cite{cahill2014nanoscale,cahill2003nanoscale} have illustrated the importance of accounting for the phonon dispersion and polarization.
For example, the average mean free path of phonons in silicon is approximate 300 nm when phonon dispersion and polarization is accounted, as opposed to 43 nm if neglecting the phonon dispersion and polarization~\cite{ju1999phonon}.
Besides, the phonon scattering and information exchange between different frequencies have great influence on the thermal conductivity of materials in different temperature range and scales~\cite{chung2004role,pop2004analytic,holland1963analysis,terris2009modeling}.
Therefore, in order to capture the real heat transfer physics in materials more accurately, the non-gray model including the phonon dispersion and polarization~\cite{MurthyJY05Review} is more preferable compared to the other two models.

Over the past decades, a lot of numerical methods are developed to solve the phonon BTE based on the non-gray model, such as the (control angle) discrete ordinate method (DOM, CADOM)~\cite{ChaiJC93RayEffect,SyedAA14LargeScale}, hybrid Ballistic-Diffusive or Fourier-BTE method~\cite{Pareekshith16BallisticDiffusive,MurthyJY12HybridFBTE}, Monte Carlo (MC) method~\cite{PJP11MC,Hadjiconstantinou15MC,MazumderS01MC,Lacroix05} and so on.
The Monte Carlo method~\cite{MazumderS01MC,mittal2010monte,Lacroix05,randrianalisoa2008monte,HEATGENERATION11} is one of the most widely used statistics methods and has made great progress in thermal application.
But its time step and grid size have to be smaller than the relaxation time and the phonon mean free path.
Moreover, it suffers large statistics errors in the diffusive regime and becomes prohibitively expensive.
Some improvements of the Monte Carlo method have also been made~\cite{PJP11MC,Hadjiconstantinou15MC} to mitigate the statistics errors.
Different from the Monte Carlo method, the DOM (CADOM)~\cite{ChaiJC93RayEffect,SyedAA14LargeScale} solves the phonon BTE directly.
It discretizes the whole wave vector space into a lot of small pieces to capture the highly non-equilibrium distribution function, which increases memory requirement significantly by several orders of magnitude.
The implicit DOM with sequential solution algorithm~\cite{terris2009modeling,wangmr17callaway,narumanchi2005comparison,hsieh2012thermal} is developed for steady problems.
It accelerates convergence when the Knudsen number (Kn, the ratio of the phonon mean free path to the characteristic length of the system) is large, but has slow convergence as the Knudsen number decreases~\cite{MurthyJY15COMET,narumanchi2005comparison}.
In order to solve this problem, the coupled ordinate method (COMET)~\cite{MurthyJY15COMET} adopts the fully implicit scheme to treat the scattering term so that it can describe the thermal transport mechanisms in different scales efficiently.
But to solve the phonon BTEs in the whole wave vector space simultaneously will generate a huge matrix, which may be very complex and difficult to solve.
For the hybrid methods~\cite{Pareekshith16BallisticDiffusive,MurthyJY12HybridFBTE}, a cutoff Knudsen number is introduced to determine whether the (implicit) DOM or approximate diffusive equation is adopted in a portion of phonon frequencies and branches.
It predicts well both in the ballistic and diffusive regimes.
However, the choice of a reasonable cutoff Knudsen number is still an open question and affects the numerical solution significantly.

Recently, we proposed an efficient implicit kinetic scheme~\cite{Chuang17gray}, which stemmed from the rarefied gas dynamics~\cite{ChenSz17Memory,3DZhu17}, to solve the multiscale problem of the gray model for all Knudsen numbers.
The core of this scheme is the introduction of the macroscopic iteration which is free of any artificial parameter like cutoff Knudsen number. This novel implicit scheme does not store any microscopic distribution function, only requires the same amount of the memory as the Fourier solver does. 

In our previous work, the phonon dispersion and polarization are not considered, which significantly limits its application in real thermal engineering~\cite{chung2004role,pop2004analytic,holland1963analysis,terris2009modeling,Minnich15advances,cahill2014nanoscale,cahill2003nanoscale}.
If including the phonon dispersion and polarization, for example adopting the non-gray model, the numerical simulations will increase two degrees of freedom, which greatly increases the computational cost.
Besides, in room temperature silicon the phonon mean free path may range over 4–5 orders of magnitude, which indicates that the phonon BTE in the whole frequency space is essentially multiscale.
Hence, it is a challenging and meaningful work to develop an extension of our previous work from gray model to non-gray model.

In this study,  an efficient implicit kinetic scheme is developed to solve the stationary phonon BTE based on the non-gray model.
The rest of this article is organized as follows. In Sec. 2, the phonon BTE based on the non-gray model and dispersion relations are introduced, and in Sec. 3, the whole algorithm of the present scheme and the boundary conditions are introduced and discussed in detail; in Sec. 4, the performance of the present scheme is tested through three numerical tests; finally, a conclusion is drawn in the last section.

\section{Theory and mathematical formula}

\subsection{Boltzmann transport equation}
The stationary phonon Boltzmann transport equation (BTE) with the single-mode relaxation time approximation~\cite{srivastava1990physics,ZimanJM60phonons,ChenG05Oxford} can be written as
\begin{equation}
\bm {v} \cdot \nabla{f}=\frac{f^{eq}-f}{\tau}, \label{eq:BTEf}
\end{equation}
where $f$ is the desired distribution function of phonons, $\bm v$ is the group velocity, $\tau$ is the effective relaxation time, $f^{eq}$ is the associated desired equilibrium distribution function.
The desired distribution function $f=f(\bm{x},\bm{s},\omega,p)$, is a function of the space vector $\bm{x}$ (three components), directional unit vector $\bm{s}$ ($\bm{s}=\bm{s}(\theta, \varphi)$, two components: the polar angle $\theta$ and the azimuthal angle $\varphi$), angular frequency $\omega$ and polarization $p$.
The angular frequency $\omega$ is given through the phonon dispersion relation $\omega=\omega(\bm{K},p)$, where $\bm{K}$ is the whole wave vector space, which is assumed to be isotropic.
The group velocity $\bm{v}$ can be calculated by $\bm{v}=\nabla_{\bm{K}}{\omega}$.
The equilibrium distribution function, i.e., $f^{eq}$,  follows the Bose-Einstein distribution,
\begin{equation}
f^{eq}=\frac {1}{\exp(\hbar\omega/k_{B}T)-1},
\label{eq:Bosefeq}
\end{equation}
where $\hbar$ is the Planck's constant divided by $2\pi$, $k_B$ is the Boltzmann constant, $T$ is the temperature.
The effective relaxation time ($\tau=\tau(\omega,~p,~T)$), which is a combination of all scattering processes~\cite{narumanchi2004submicron,holland1963analysis,armstrong1985n}, such as the boundary scattering , impurity scattering, umklapp (U) and normal (N) phonon-phonon scattering etc, and can be estimated using the Matthiessen's rule~\cite{MurthyJY05Review}
\begin{equation}
\frac{1}{\tau}=\frac{1}{\tau_{\text{boundary}}}+\frac{1}{\tau_{\text{impurity}}}+\frac{1}{\tau_{\text{U}}}+\frac{1}{\tau_{\text{N}}}+ ...
\label{eq:Matthiessen}
\end{equation}

Generally, the Equation~\eqref{eq:BTEf} can be written as follows,
\begin{equation}
\bm v \cdot \nabla e  = \frac { e^{eq}-e }{\tau},
\label{eq:BTEe}
\end{equation}
where
\begin{align}
e(\bm{x},\bm{s},\omega,p) &=\hbar \omega (f-f^{eq}(T_{\text{ref}})) D(\omega, p)/4{\pi},  \\
e^{eq}(\bm{x},\omega,p) &=\hbar \omega (f^{eq}-f^{eq}(T_{\text{ref}})) D(\omega, p)/4{\pi},
\end{align}
where $e$ is the desired energy density distribution function, $e^{eq}$ is the associated desired equilibrium energy density distribution function, $D(\omega,p)$ is the phonon density of states and $T_{\text{ref}}$ is the reference temperature.
If integrate Eq.~\eqref{eq:BTEe} over the whole solid angle space ($\Omega$) and the frequency space ($\omega,~p$), it can be obtained
\begin{equation}
\nabla \cdot \bm{q}  = \sum_{p}\int_{\omega_{min,p}}^{\omega_{max}} \int_{4\pi} \frac { e^{eq}-e }{\tau} d{\Omega}d{\omega},
\label{eq:energyconservation}
\end{equation}
where
\begin{equation}
\bm{q} =\sum_{p}\int_{\omega_{min,p}}^{\omega_{max,p}} \int_{4\pi} \bm v  e d{\Omega}d{\omega}
\label{eq:heatflux}
\end{equation}
is the total heat flux, $\omega_{min,p}$ and $\omega_{max,p}$ are the minimum and maximum angular frequency corresponding to a given phonon polarization branch $p$.
At steady state, the divergence of the heat flux must be null, therefore, it can be obtained
\begin{equation}
\sum_{p}\int_{\omega_{min,p}}^{\omega_{max,p}} \int_{4\pi} \frac { e^{eq}-e }{\tau} d{\Omega}d{\omega}=0.
\label{eq:scatteringterm}
\end{equation}

The temperature differences ($DT$) in the whole domain are assumed to be small enough compared to the reference temperature ($T_{\text{ref}}$) of the system, i.e., $DT \ll T_{\text{ref}}$.
Then the following approximation can be made~\cite{MurthyJY12HybridFBTE,narumanchi2004submicron}
\begin{equation}
\int_{4\pi} e^{eq} d{\Omega}\approx C (T-T_{\text{ref}}),
\label{eq:specificheat1}
\end{equation}
where $C=C(\omega,p,T)$ is the mode specific heat, defined as
\begin{equation}
C =\hbar \omega D(\omega,p) \frac{\partial{f^{eq}}}{\partial{T}},
\label{eq:specificheat}
\end{equation}
where $D(\omega,p)= {k^2}/({2\pi^{2}|\bm{v}|})$.
The relaxation time ($\tau$) and the specific heat ($C$) discussed in the following are both calculated based on the reference temperature ($T_{\text{ref}}$), i.e., $\tau=\tau(\omega,p,T_{\text{ref}})$, $C=C(\omega,p,T_{\text{ref}})$.
Then, the Eq.~\eqref{eq:scatteringterm} can be written as follows,
\begin{equation}
\sum_{p}\int_{\omega_{min,p}}^{\omega_{max,p}} \frac{C \left( T-T_{\text{ref}} \right) -  \int_{4\pi} e d\Omega } {\tau } d{\omega} = 0.
\label{eq:scatteringcp}
\end{equation}
The temperature of the system can be calculated based on the Eq.~\eqref{eq:scatteringcp}, i.e.,
\begin{equation}
T=T_{\text{ref}}+ \left( \sum_{p}\int_{\omega_{min,p}}^{\omega_{max,p}}  \frac{\int_{4\pi} e d\Omega }{\tau } d{\omega}   \right)/ \left(  \sum_{p}\int_{\omega_{min,p}}^{\omega_{max,p}}  \frac{ C}{\tau }  d{\omega} \right).
\label{eq:calT}
\end{equation}
It is important to note that the temperature calculated in the above procedures is not the thermodynamics temperature, but a reasonable description of the local internal energy~\cite{MurthyJY12HybridFBTE,narumanchi2004submicron,SyedAA14LargeScale,ChenG05Oxford}.
For the sake of convenience, it is referred to as the temperature in the following discussions.
%

\subsection{Dispersion relations and basic parameters}

In the present work, let's take the monocrystalline silicon as an example.
The phonon dispersion relation of the silicon in the [100] direction are chosen to represent the other directions~\cite{holland1963analysis,brockhouse1959lattice}.
The silicon has three acoustic phonon branches (one longitudinal acoustic phonon branch (LA) and two degenerate transverse acoustic phonon branches (TA)) and three optical phonon branches (one longitudinal optical phonon branch (LO) and two degenerate transverse optical phonon branches (TO)).
Only the acoustic (LA and TA) phonon branches are considered because the optical phonon branches contribute little to the thermal conduction.
These curves of the acoustic phonon branches are fitted by the quadratic polynomial dispersions, which can be expressed as
\begin{equation}
\omega=c_0+c_{1}k+c_{2}k^2,
\label{eq:curves}
\end{equation}
where the wave vector $k \in[0,k_{max}]$, $k_{max}=2\pi /a $ is the maximum wave vector in the first Brillouin zone, $a$ is the lattice constant (for silicon, $a=5.43$\r{A}).
In this study, the Pop's dispersions~\cite{pop2004analytic} are adopted. (For LA, $c_0=0$ \text{rad/s}, $c_1=9.01\times 10^{-3}$ \text{m/s}, $c_2=-2.0 \times 10^{-7}$ ${\text{m}}^{\text{2}}/{\text{s}}$, while for TA, $c_0=0$ \text{rad/s}, $c_1=5.23\times 10^{-3}$ \text{m/s}, $c_2=-2.26 \times 10^{-7}$ ${\text{m}}^{\text{2}}/{\text{s}}$.)

The group velocity can be obtained through the dispersion relations (Eq.~\eqref{eq:curves}),
\begin{align}
\bm{v} &=|\bm{v}|\bm{s}=(c_{1}+2c_{2}k)\bm{s}.
\end{align}
While for the relaxation time, the Terris' relaxation time~\cite{terris2009modeling} is used.
($\tau^{-1}=\tau_{{\text{impurity}}}^{-1}+\tau_{{\text{U}}}^{-1}+\tau_{{\text{N}}}^{-1}$;
$\tau_{{\text{impurity}}}^{-1}=A_{i}\omega^{4}$,
where $A_{i}=1.498\times10^{-45}~{\text{s}^{\text{3}}}$;
for LA, $\tau_{{\text{NU}}}^{-1}=\tau_{{\text{N}}}^{-1}+\tau_{{\text{U}}}^{-1}=B_{L}\omega^{2}T^{3}$,
where $B_{L}=1.180\times 10^{-24}~{\text{K}^{\text{-3}}}$;
for TA, if $0 \leq k < k_{max}/2$,
$\tau_{{\text{N}}}^{-1}=B_T\omega T^4$,
if $k_{max}/2 \leq k \leq k_{max}$,
$\tau_{{\text{U}}}^{-1}=B_U\omega^{2}/{\sinh(\hbar\omega/k_{B}T)}$,
where $B_T=8.708\times 10^{-13}~{\text{K}^{\text{-3}}}$,
$B_{U}=2.890\times10^{-18}~{\text{s}}$.)
Once the relaxation time and the group velocity are known, the Knudsen number ($\text{Kn}$) can be obtained through its definition: $\text{Kn}=\lambda/L_{\text{ref}}$, where $\lambda=|\bm{v}|\tau$ is the phonon mean free path, $L_{\text{ref}}$ is the characteristic length of the system.


\section{Numerical scheme}

As mentioned in the previous subsection, the phonons with different frequencies and branches may transport in different regimes.
In order to accurately predict the multiscale heat transfer phenomena, an implicit kinetic scheme based on the non-gray model is proposed to capture the thermal transport physics on different scales simultaneously.
The proposed scheme is composed of two parts, the microscopic iteration and the macroscopic iteration, which are introduced in this section.

\subsection{Microscopic iteration}

The microscopic iteration solely solves the phonon BTE within a step, and provides the distribution functions and their moments (i.e., the macroscopic heat flux ($\bm{q}$) and the intermediate temperature (${T}^{**}$)) for the subsequent macroscopic iteration.

The wave vector space is discretized as follows.
For each phonon branch ($p$), the wave vector $k$ is discretized equally into $N_B$ discrete bands ($k_b=k_{max}(2b-1)/(2N_B)$, where the index of the discretized bands $b$ satisfies $b \in[1,N_B]$).
The corresponding discretized angular frequency can be obtained through the dispersion relations $\omega_b=\omega(k_b,p)$.
Besides, the basic parameters (group velocity $\bm{v}$, relaxation time $\tau$ and specific heat $C$) needed in the phonon BTE solver are calculated based on the reference temperature ($T_{\text{ref}}$), corresponding discretized angular frequency ($\omega$) and phonon branch ($p$).
The numerical integration over the frequency space is performed using the midpoint rectangular rules.
While for the discretization of the solid angle space ($\Omega$) or the direction ($\bm{s}=\left( \cos \theta, \sin \theta \cos \varphi, \sin \theta \sin \varphi \right)$), in 1D cases, the Gauss-Legendre quadrature~\cite{NicholasH13GaussL} with $N_{\cos \theta}$ points is employed to discrete the $\cos {\theta}$ in $[-1,1]$.
In 2D cases, the CADOM~\cite{SyedAA14LargeScale,MurthyJY98FVMradiative} is used to enhance the numerical conservation.
The polar angle $\theta$ ($\theta \in [0, \pi]$) and azimuthal angle $\varphi$ ($\varphi \in [0,\pi]$ due to symmetry) are discretized equally into $N_{\theta}$ and $N_{\varphi}$ discretized angles, respectively.

Under the discretized wave vector space, the governing equation of the desired energy density distribution function ($e$) at steady state, namely Eq.~\eqref{eq:BTEe}, can be written as follows,
\begin{equation}
|\bm v_{\omega,p}| \bm s_{\alpha} \cdot \nabla e_{\omega,p,\alpha}  = \frac { e^{eq}_{\omega,p}-e_{\omega,p,\alpha} }{\tau_{\omega,p}},
\label{eq:BTEd}
\end{equation}
where the subscript index $\omega,~p,~\alpha$ means the discretized phonon frequency, polarization and direction, respectively.
An iterative scheme is constructed to solve the stationary phonon BTE,
\begin{equation}
|\bm v_{\omega,p}| \bm s_{\alpha} \cdot \nabla  e_{\omega,p,\alpha}^{n+1}  = \frac { e^{eq}_{\omega,p}(T^n)  -e_{\omega,p,\alpha}^{n+1} }{\tau_{\omega,p}},
\label{eq:BTEdn}
\end{equation}
where $n$ is the iteration number.
In order to solve $e_{\omega,p,\alpha}^{n+1}$ in the above equation, an inner iteration is introduced as follows,
\begin{equation}
 \frac { \Delta e_{\omega,p,\alpha}^{n,m+1} }{\tau_{\omega,p}}+|\bm v_{\omega,p}| \bm s_{\alpha} \cdot \nabla (\Delta e_{\omega,p,\alpha}^{n,m+1} ) = \text{res}_{\omega,p,\alpha}^{n,m},
\label{eq:BTEdelta}
\end{equation}
where $m$ is the micro inner iteration number, $\Delta e_{\omega,p,\alpha}^{n,m+1} =e_{\omega,p,\alpha}^{n,m+1}- e_{\omega,p,\alpha}^{n,m}$, $\text{res}_{\omega,p,\alpha}^{n,m}$ is the residual of the micro inner iteration, defined as
\begin{equation}
\text{res}_{\omega,p,\alpha}^{n,m}= \frac { e^{eq}_{\omega,p}(T^n)  -e_{\omega,p,\alpha}^{n,m} }{\tau_{\omega,p}} - |\bm v_{\omega,p}| \bm s_{\alpha} \cdot \nabla  e_{\omega,p,\alpha}^{n,m}.
\label{eq:microres}
\end{equation}
Let $e_{\omega,p,\alpha}^{n,0}= e^{eq}_{\omega,p}(T^n)$ to initiate the inner iteration.

The left hand side of the Eq.~\eqref{eq:BTEdelta} is approximated by the first-order upwind scheme, which is numerically stable and easy for coding.
While the residual of the micro inner iteration can be calculated by arbitrary order numerical schemes based on demand.
Then the Eq.~\eqref{eq:BTEdelta} in integral form over a control volume becomes
\begin{equation}
\begin{aligned}
&\left( \frac{V_i}{\tau_{\omega,p}} + \frac{1}{2} |\bm v_{\omega,p}|\sum_{j \in N(i)} |\mathbf{n}_{ij} \cdot \bm{s}_{\alpha} | S_{ij}  \right) \Delta e_{\omega,p,\alpha,i}^{n,m+1} +   \frac{1}{2}|\bm v_{\omega,p}| \sum_{j \in N(i)}  \left( \mathbf{n}_{ij} \cdot \bm{s}_{\alpha}- |\mathbf{n}_{ij} \cdot \bm{s}_{\alpha} | \right)  S_{ij} \Delta e_{\omega,p,\alpha,i}^{n,m+1} \\
&=\text{res}_{\omega,p,\alpha,i}^{n,m} = \frac{V_i}{\tau_{\omega,p}} \left(  e^{eq}_{\omega,p}(T_{i}^n)  -e_{\omega,p,\alpha,i}^{n,m} \right)  - |\bm v_{\omega,p}| \sum_{j\in N(i)} \mathbf{n}_{ij} \cdot \bm{s}_{\alpha} { e_{\omega,p,\alpha,ij}^{n,m}} S_{ij},
\end{aligned}
\label{eq:BTEcontrol}
\end{equation}
where $V_i$ is the volume of the cell $i$, $N(i)$ denotes the sets of face neighbor cell of cell $i$, $ij$ denotes the interface between cell $i$ and cell $j$, $S_{ij}$ is the area of the  interface $ij$, $\mathbf{n}_{ij}$ is the normal of the interface $ij$ directing from cell $i$ to cell $j$.
The Lower-Upper Symmetric Gauss-Seidel scheme (LU SGS)~\cite{YoonS88LUSGS} is adopted to solve the Eq.~\eqref{eq:BTEcontrol}.
A non-dimensional and frequency-dependent error ($\epsilon_1$) is introduced and defined as
\begin{equation}
\epsilon_1= \text{max} \left\{ \left| \frac{  \tau_{\omega,p} \text{res}_{\omega,p,\alpha,i}^{n,m}  } {  V_{i} C_{\omega,p} \times DT \times \text{Kn}_{\omega,p} }  \right| \right\},
\label{eq:epsilon1}
\end{equation}
where $\text{max}$ means the maximum value over all cells, $\text{Kn}_{\omega,p}=|\bm{v}_{\omega,p}| \tau_{\omega,p} /L_{\text{ref}}$.
When $\epsilon_1 $ is smaller than a given threshold, the micro inner iteration converges.
Let $e_{\omega,p,\alpha}^{n+1}= e_{\omega,p,\alpha}^{n,M_0}$, where $M_0$ is the iterative number required for the converged micro inner iteration.
Add its contribution to a temporary variable immediately for the update of the intermediate temperature ($T^{**}$) and the heat flux ($\bm{q}^{n+1}$) at the next iteration step, i.e.,
\begin{equation}
T^{**}=T_{\text{ref}}+  \frac{ \sum_{p} \sum_{\omega} \sum_{\alpha} \left( w_{\omega} w_{\alpha}  e_{\omega,p,\alpha}^{n+1}  / \tau_{\omega,p} \right)  }{ \sum_{p} \sum_{\omega}  \left( C_{\omega,p}/\tau_{\omega,p} \right)  } ,
\label{eq:calT2}
\end{equation}
\begin{equation}
\bm{q}^{n+1} =\sum_{p} \sum_{\omega} \sum_{\alpha} w_{\omega} w_{\alpha} |\bm v_{\omega,p}|\bm{s}_{\alpha}  e_{\omega,p,\alpha}^{n+1} ,
\label{eq:heatflux2}
\end{equation}
where $w_{\omega},~w_{\alpha}$ are the associated weights for the discretized frequency space and solid angle space (direction), respectively.

The present microscopic iteration, is similar to the implicit DOM~\cite{terris2009modeling,wangmr17callaway,narumanchi2005comparison,hsieh2012thermal}, which converges slowly as the Knudsen number decreases~\cite{MurthyJY15COMET,narumanchi2005comparison,MurthyJY12HybridFBTE}.
In order to ensure the high convergence efficiency for all phonons transporting with various frequencies, the macroscopic iteration is introduced on the basis of the microscopic iteration.


\subsection{Macroscopic iteration}

The corresponding macroscopic governing equation of the Eq.~\eqref{eq:BTEe} is
\begin{equation}
\nabla \cdot \bm{q} =0,
\label{eq:macroequation}
\end{equation}
where the heat flux $\bm{q}=\bm{q}(T)$, which can be regarded as a functional of the temperature field $T(\bm{x})$ on a macroscopic view. Although no universal formula for the heat flux can be established at the micro/nano scale,
the heat flux ($\bm{q}$) can be explicitly calculated by the moments of the distribution function (Eq.~\eqref{eq:heatflux}) in the framework of the phonon BTE.
In the last subsection, the numerical heat flux is obtained by the Eq.~\eqref{eq:heatflux2} in the microscopic iteration.
Then a macroscopic residual ($\text{RES}$) is defined as
\begin{equation}
\text{RES}^n = \text{RES}(T^n) = - \nabla \cdot \bm{q}^{n+1}.
\label{eq:macrores}
\end{equation}
An approximate linear operator ($\tilde Q$)~\cite{Chuang17gray,RonSD82Newton,Andrew78Newton,Moore68Newton} is invoked and acts on the increment of the temperature ($\Delta T$),
\begin{equation}
\tilde {Q} (\Delta T^{n}) = \text{RES}^n,
\label{eq:macroiteration}
\end{equation}
so that $||\text{RES}(T^n+\Delta T^{n})|| < ||\text{RES}(T^n)||$, where $\Delta T^{n}=T^{n+1}-T^{**} $, $T^{**}$ is obtained by Eq.~\eqref{eq:calT2}.
As the residual goes to zero, Eq.~\eqref{eq:macroequation} can be satisfied.

The formula of the approximate linear operator is
\begin{equation}
\tilde {Q} (\Delta T) =  \nabla \cdot \left( -\beta k_{{bulk}} \nabla \left(\Delta T\right) \right),
\label{eq:similarfourier}
\end{equation}
where $\beta$ is a non-dimensional coefficient which can be adjusted to ensure the convergence of the iteration, and
\begin{equation}
k_{{bulk}}=\frac{1}{3} \sum_{p} \int_{\omega_{min,p}}^{\omega_{max,p}} C|\bm{v}|^{2} \tau d{\omega}
\label{eq:bulkconductivity}
\end{equation}
is the bulk thermal conductivity obtained in the diffusive limit.
Based on the theorem of the inexact Newton method~\cite{RonSD82Newton,Andrew78Newton,Moore68Newton}, the iteration may converge within a certain range of $\beta$.
As long as the iteration converges, the approximate linear operator will not influence the final convergent solution.
Without special statements, $\beta$ is set to be a constant in the whole iterative process for simplicity.

Combining Eq.~\eqref{eq:macroiteration} and Eq.~\eqref{eq:similarfourier}, the finite volume discretization of  Eq.~\eqref{eq:macroiteration} can be written as
\begin{equation}
- \frac{k_{{bulk}}}{V_i} \sum_{j \in N(i)} S_{ij} \mathbf{n}_{ij} \cdot \left[ \beta \nabla  \left( \Delta T_{ij}^{n}  \right) \right]  = \text{RES}_{i}^{n},
\label{eq:FVMiteration}
\end{equation}
where
\begin{equation}
\text{RES}_{i}^{n}= - \frac{1}{V_i}  \sum_{j \in N(i)} S_{ij} \mathbf{n}_{ij} \cdot \bm q_{ij}^{n+1} = - \frac{1}{V_i}  \sum_{j \in N(i)} S_{ij} \mathbf{n}_{ij} \cdot  \left(  \sum_{p} \sum_{\omega} \sum_{\alpha} w_{\omega} w_{\alpha} |\bm v_{\omega,p}|\bm{s}_{\alpha}  e_{ij,\omega,p,\alpha}^{n+1}  \right).
\label{eq:macroRES}
\end{equation}
The equation~\eqref{eq:FVMiteration} in 1D and 2D cases is solved by the Thomas algorithm~\cite{datta2010numerical} and the conjugate gradient (CG) method~\cite{ADAMS02fastiterative}, respectively.
Because the macroscopic residual (${\text{RES}}$) is obtained through the microscopic phonon distribution function, the numerical accuracy of the macroscopic iteration (or the present implicit kinetic scheme) is totally controlled by the microscopic iteration.

After the macroscopic iteration, the temperature field and the associated equilibrium state of the next microscopic iteration can be updated.
It is obvious that the computational cost of the macroscopic iteration basically can be ignored compared to the microscopic iteration, because the phase space of macroscopic equation is far less than that of the phonon BTE.

\subsection{Boundary conditions}

The boundary conditions play an indispensable role in the simulation.
For the phonon BTE, the isothermal boundary condition, diffusely reflecting boundary condition and periodic boundary condition are presented as follows.

\begin{enumerate}
  \item The isothermal, thermalizing boundary assumes that the incident phonons are all absorbed by the boundary ($\bm{x}_{b}$), while the phonons emitted from the boundary are the equilibrium state with the boundary temperature ($T_{b}$).
      The mathematical formula can be expressed as
      \begin{equation}
      e(\bm{x}_{b},\bm{s},\omega,p)=e^{eq}(T_{b},\omega,p), \quad \bm{s} \cdot \mathbf{n}_{b} >0,
      \label{eq:BC1}
      \end{equation}
      where $\mathbf{n}_{b}$ is the normal unit vector of the boundary pointing to the computational domain.
  \item The diffusely reflecting boundary condition is a kind of the adiabatic boundary.
      It assumes that the phonons reflected from the boundary are equal along each direction and the net heat flux across the boundary is zero.
      Mathematically, it can be written as
      \begin{equation}
      e( \bm{x}_{b}, \bm{s}, \omega,p)= \frac{1}{\pi}  \int_{ \bm{s}' \cdot \mathbf{n}_{b} < 0}{ e(\bm{x}_{b},\bm{s}',\omega,p)  \left| \bm{s}' \cdot  \mathbf{n}_{b}  \right| }d\Omega, \quad \bm{s} \cdot \mathbf{n}_{b} >0.
      \label{eq:BC2}
      \end{equation}
   \item In the periodic boundary, when a phonon leaves the computational domain from one periodic boundary, another phonon with the same group velocity and frequency will enter the computational domain from the corresponding periodic boundary at the same time.
      Besides, the deviations from the local equilibrium states of the distribution functions of the two phonons are the same, i.e.,
      \begin{equation}
      e(\bm{x}_{b_1}, \bm{s}, \omega,p )- e^{eq} ({T}_{b_1}, \omega,p ) = e(\bm{x}_{b_2}, \bm{s}, \omega,p )- e^{eq} ({T}_{b_2}, \omega,p ),
      \label{eq:BC3}
      \end{equation}
      where $\bm{x}_{b_1}$, ${T}_{b_1}$ and $\bm{x}_{b_2}$, ${T}_{b_2}$ are the space vector and temperature of the two associated periodic boundary $b_1$ and $b_2$, respectively.
\end{enumerate}

The boundary conditions are only employed to evaluate the residual of the micro inner iteration ($\text{res}$) and the macroscopic residuals ($\text{RES}$), does not affect the iteration procedure.
The ghost cells ($gc$) are used in the process of solving Eq.~(\ref{eq:BTEcontrol},~\ref{eq:FVMiteration}).
When solving them, let $\Delta T_{gc}\equiv 0 $ and $\Delta e_{gc} \equiv 0 $ respectively regardless of the boundary condition, because the increment of any quantity finally vanishes.

%

\section{Numerical tests}

To validate the performance of the present scheme, three numerical tests are simulated, including the 1D cross-plane heat conduction~\cite{sellan2010cross,MajumdarA93Film}, 2D in-plane~\cite{turney2010plane,Cuffe15conductivity} and square-cylinder nanoporous~\cite{Bera10nanoporous,hsieh2012thermal,LIANG20174915} heat transfer problem.

Without otherwise statements, the reference temperature of the computational system is set to be $T_{\text{ref}}=300\text{K}$ and the first-order upwind scheme is adopted to calculate the residual of the micro inner iteration in Eq.~\eqref{eq:BTEcontrol}.
In addition, a non-dimensional error ($\epsilon_2$) is introduced and defined as
\begin{equation}
\epsilon_2=\frac{ \sqrt {\sum_{i=1}^{N_{cell}}{(T_i^{n}-T_i^{n+1})^2}}  } { \sqrt {{N_{cell}}{(DT \times DT)}} },
\label{eq:epsilon2}
\end{equation}
where $N_{cell}$ is the number of the total cells in the computational domain.
When $\epsilon_2$ is smaller than a given threshold, the iteration converges.
To find the optimized number of frequency bands, the bulk thermal conductivity is calculated with $N_B = 10, 20, 40, 100$.
The corresponding bulk thermal conductivity is $146.0$ W/(m.K), $145.9$ W/(m.K), $145.9$ W/(m.K) and $145.9$ W/(m.K), respectively.
Therefore, when $N_B \geq 20$, the numerical integration in the frequency space is regarded as converged.

\subsection{1D cross-plane heat conduction}

Cross-plane heat conduction phenomena are very common in the experiments~\cite{sellan2010cross,MajumdarA93Film,hua2015semi}, for example the measurement of the thermal conductivity of a film.
The thickness of the film is $L$ ($L_{\text{ref}} =L$).
The temperature of the left boundary of the film is fixed at $T_{L}=T_{\text{ref}}+DT/2$, while the right is $T_{R}=T_{\text{ref}}-DT/2$.
The thermailizing boundary condition is implemented on both two boundaries.
The whole computational domain is discretized equally with $N_{cell}$ cells.
One hundred ($N_{\cos \theta} =100$) points are employed for the discretization of the directions to ensure the numerical accuracy at different scales.
Set $N_B=40$, $N_{cell}=40$, and $\beta=L/L_{0}$  (Eq.~\eqref{eq:similarfourier}), where $L_{0}=1$ um, and the vanLeer limiter is adopted to deal with the residual of the micro inner iteration in Eq.~\eqref{eq:BTEcontrol} to enhance the numerical accuracy and stability.
When $\epsilon_2 < 1.0 \times 10^{-9}$, the iteration converges.

\begin{figure}
 \centering
 \includegraphics[width=0.45\textwidth]{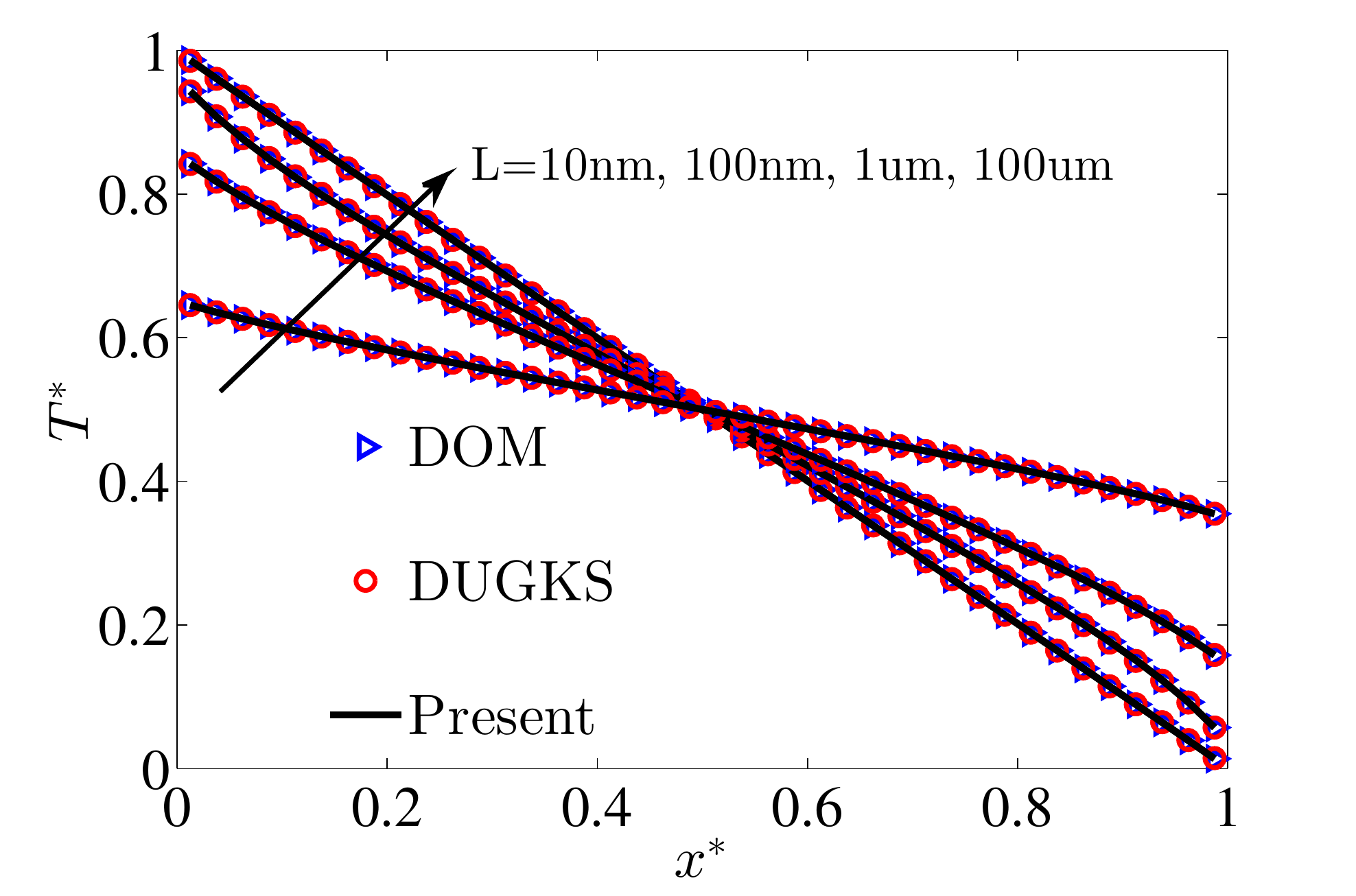}~
 \caption{
 Non-dimensional temperature field with the different thickness ($L=$ 10nm, 100nm, 1um, 100um) of the film, where $T^{*}=(T-T_R)/(T_L-T_R),~x^{*}=x/L$. The blue triangle is the data predicted by the implicit DOM, the red circle is the results calculated by the DUGKS~\cite{GuoZl16DUGKS,LUO2017970} and the black solid line is the present numerical results.
 }
 \label{Tfilm}
\end{figure}

\begin{figure}
 \centering
 \includegraphics[width=0.45\textwidth]{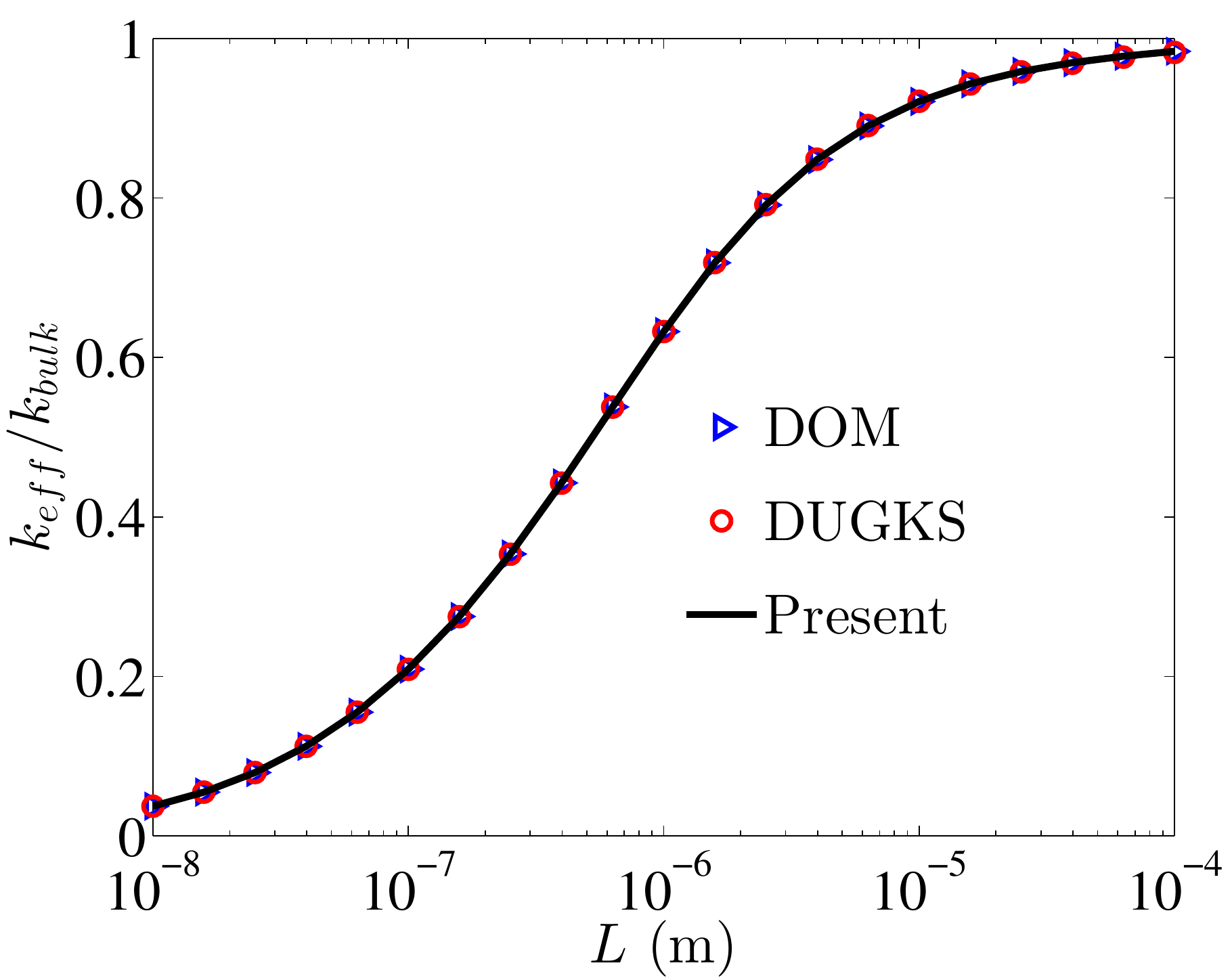}~
 \caption{
 Effective thermal conductivity, normalized by the bulk thermal conductivity ($k_{eff}/k_{bulk}$), at different length scales. The blue triangle is the data predicted by the implicit DOM, the red circle is the results calculated by the DUGKS~\cite{GuoZl16DUGKS,LUO2017970} and the black solid line is the present numerical results.
 }
 \label{kfilm}
\end{figure}

Figure~\ref{Tfilm} shows that the non-dimensional temperature field ($T^{*}=(T-T_R)/(T_L-T_R)$) with different thicknesses of the film, from which it can be observed that the present numerical results agree well with the data predicted by the implicit DOM and the explicit discrete unified gas kinetic scheme (DUGKS)~\cite{GuoZl16DUGKS,LUO2017970}.
As the thickness of the film increases large enough (i.e., $L=$ 100um), the temperature profiles become linear as predicted by the classical Fourier's law.
When the thickness of the film is very small, there is temperature slip close to the boundary because of the non-equilibrium phonon transport and the strong boundary scattering.
Figure~\ref{kfilm} shows the non-dimensional thermal conductivity ($k_{eff}/k_{bulk}$, where $k_{eff}$ is the effective thermal conductivity. It is calculated based on the classical Fourier's law, i.e., $k_{eff}=(|{q}_{steady}|L)/(|T_L-T_R|)$, where ${q}_{steady}$ is the heat flux across the film at the steady state.) at different scales, which proves that the present scheme can describe the heat transfer phenomena accurately from tens of nanometers to hundreds of microns.

In addition, a comparison of computational expense is made between the present implicit scheme and the implicit DOM, as shown in Table.~\ref{filmefficiency}.
It can be observed that with the macroscopic iteration, the present scheme accelerates convergence significantly, especially as the thickness of the film is large.
It can be found that the CPU time per iteration step of the present scheme is basically equal to that of the implicit DOM, which means the cost of the macroscopic iteration can be ignored compared to that of solving the frequency-dependent phonon BTE.
The total memory that the present scheme requires is approximately 10 kilobytes (KB), while the memory cost of the DUGKS is about 15  megabytes (MB).

\begin{figure}
 \centering
 \includegraphics[width=0.45\textwidth]{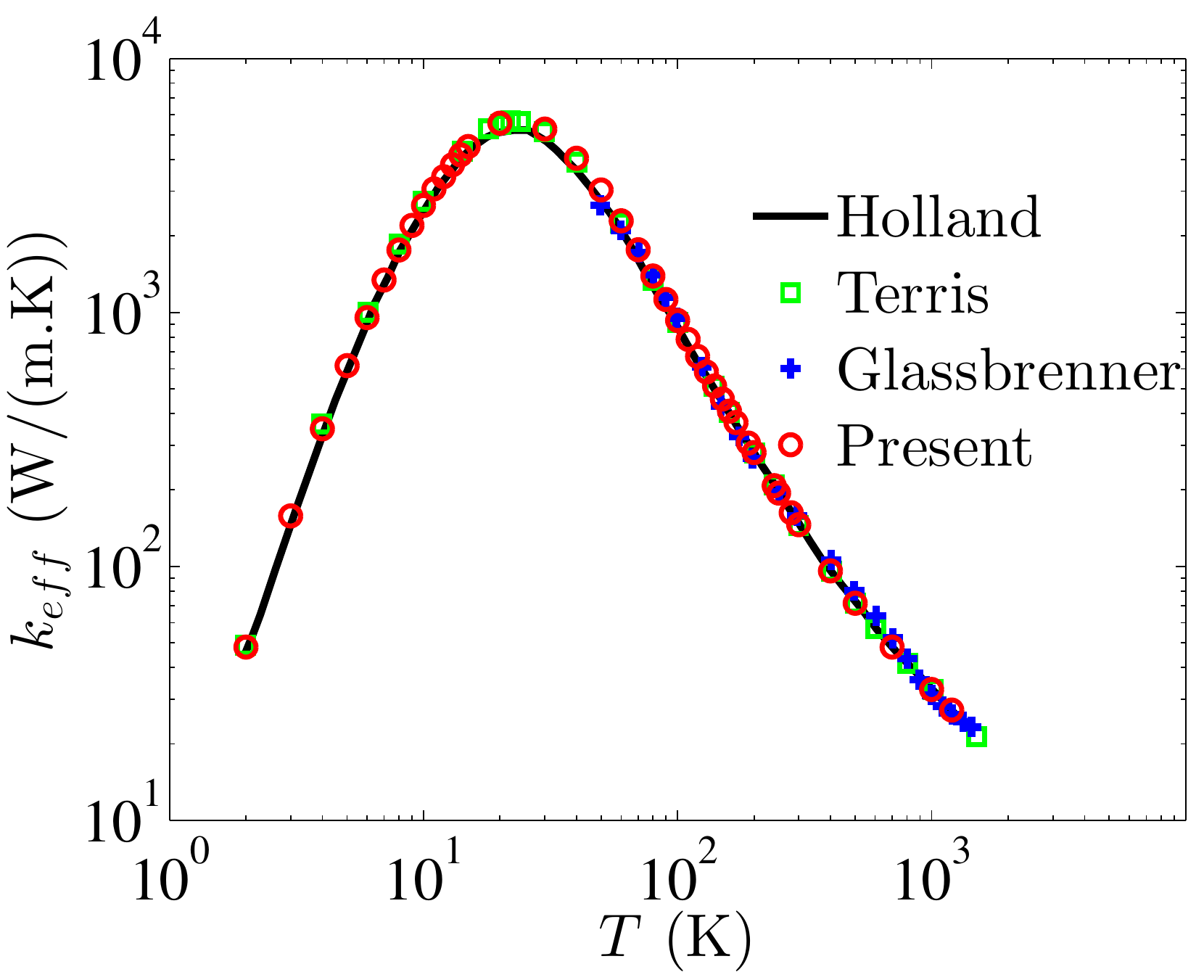}~
 \caption{
 Effective thermal conductivity of the silicon film with a thickness of 7.16 mm at different temperatures. The solid line is the thermal conductivity obtained with the Holland's method~\cite{holland1963analysis}. Green squares are the numerical results get by the Terris~\cite{terris2009modeling}. Blue cross dots (+) are the experimental data obtained by Glassbrenner and his co-workers~\cite{Glassbrenner64conductivity}. The red dots are our present results ($T_{\text{ref}} =2 {\text{K}} -1200 {\text{K}}$).
 }
 \label{filmconductivity}
\end{figure}

Above numerical results are based on the fixed reference temperature. Next, let the thickness of the film be fixed at 7.16 mm, and change the reference temperature from 2K to 1200K.
As shown in~\cref{filmconductivity}, our present numerical results are still in good agreement with the data obtained with Holland's method~\cite{holland1963analysis}, the numerical results obtained by Terris~\cite{terris2009modeling} and the experimental data obtained by Glassbrenner and his co-workers~\cite{Glassbrenner64conductivity}.
It can be observed that the present scheme can predict the silicon thermal conductivity in a wide temperature range.
Similarly, the present scheme can be used to the thermal application of other materials and nanostructures through the corresponding phonon dispersion and polarization.
\begin{table}
\caption{The efficiency of the present implicit scheme. Accelerate rate is the ratio of the CPU time between the present scheme and the implicit DOM. Steps mean the total iteration number.}
\centering
\begin{tabular}{|*{8}{c|}}
 \hline
 \multirow{2}{*}{{\shortstack{$L$ }}}  & \multicolumn{3}{c|}{Present} & \multicolumn{3}{c|} {DOM} & \multirow{2}{*}{{\shortstack{Accelerate rate}}}\\
\cline{2-7}
 & Time (s) & Steps & Time per step   & Time (s) & Steps & Time per step  &  \\
 \hline
 100 um & 33 & 1153 & 0.02862   & 22693 & 809293 & 0.02804  &  688    \\
 \hline
  10 um & 10 & 229 & 0.04367    & 631 & 15556 & 0.04056   & 63  \\
\cline{1-3}
\cline{5-6}
\cline{8-8}
 \hline
1 um & 5 & 79 & 0.06329 & 39 & 576 & 0.06732 & 8 \\
\cline{1-3}
\cline{5-6}
\cline{8-8}
 \hline
 100 nm & 2 & 22 & 0.09091 &4 & 45 & 0.08889 & 2 \\
 \hline
\end{tabular}
\label{filmefficiency}
\end{table}

\subsection{2D in-plane heat transfer}

\begin{figure}
 \centering
 \includegraphics[width=1.0\textwidth]{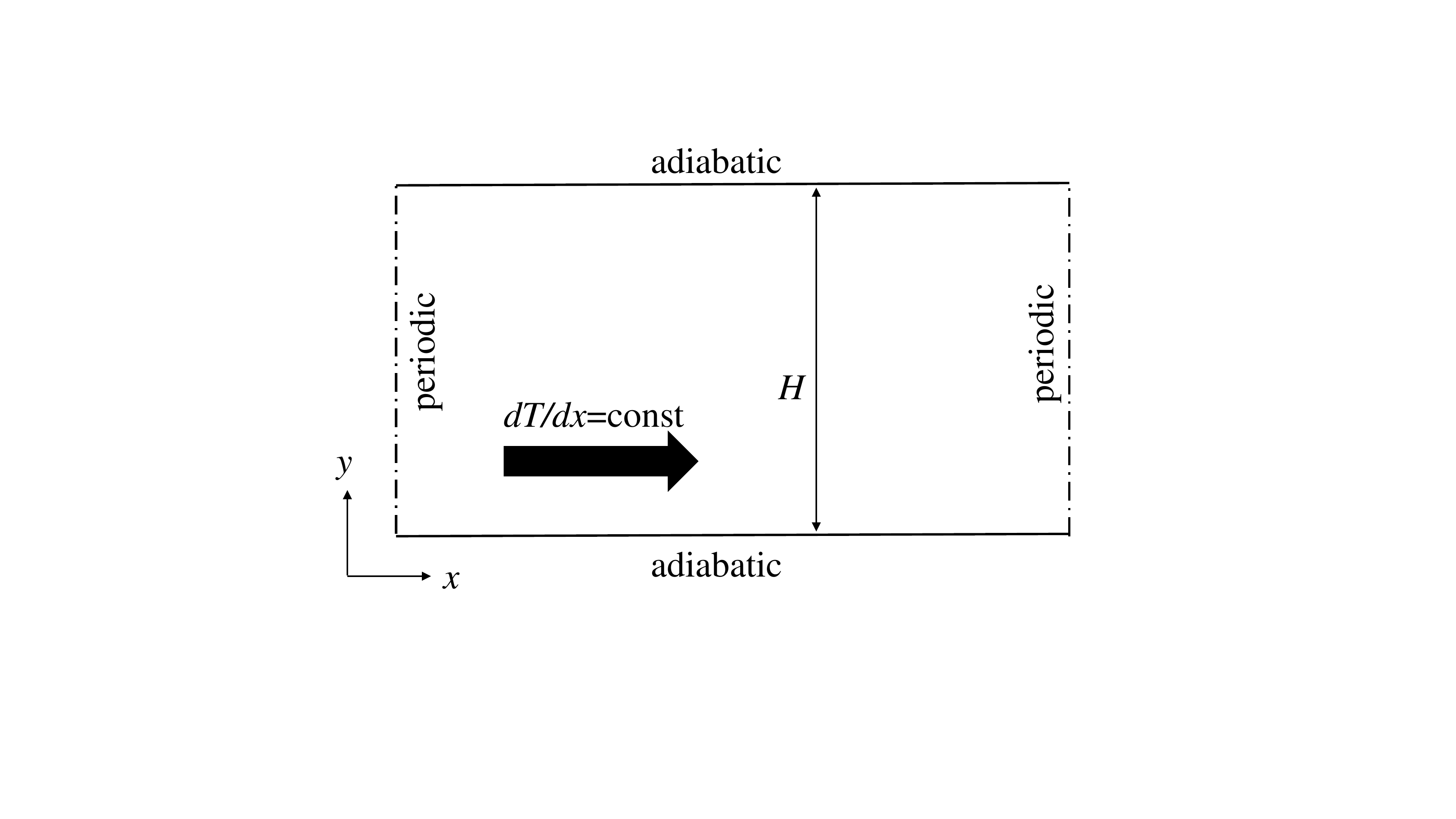}~
 \caption{Schematic of 2D in-plane heat transfer.}
 \label{inplane}
\end{figure}

The schematic of 2D in-plane heat transfer is shown in~\cref{inplane}.
Two parallel and adiabatic planes are placed with a distance of $H$ ($L_{\text{ref}} =H$) in $y$ direction.
While in $x$ direction (in-plane direction), a uniform and small enough temperature gradient ($dT/dx$) is applied.
The diffusely reflecting boundary condition is adopted for the adiabatic planes.
While the left and right boundaries are set to be periodic.

The in-plane heat transfer problem can be described by the Fuchs-Sondheimer theory and the heat flux ($\bm {q}$) in the in-plane direction can be expressed as~\cite{Cuffe15conductivity}
\begin{equation}
\begin{aligned}
q_{x}(Y)=-\frac{1}{4} \frac{dT}{dx} \sum_{p}{  \int_{0}^{\omega_{max,p}} \int_{0}^{1} C_{\omega,p} \tau_{\omega,p} |\bm{v}_{\omega,p}|^2 (1- \eta ^2)
{\times}  \left\{ 2-\exp{\left( -\frac{Y}{\eta \text{Kn}_{\omega,p}} \right)} - \exp{\left( -\frac{1-Y}{\eta \text{Kn}_{\omega,p}} \right)} \right\}   } d\eta d\omega,
\end{aligned}
\label{eq:analytical}
\end{equation}
where $Y=y/H$ is the non-dimensional ordinate.
The exact numerical solution of Eq.~\eqref{eq:analytical} is obtained by numerical integration using 4000 discrete points for $\omega$ and $\eta$ respectively.
Let the number of the cells in the $x$ direction be fixed at $N_x=20$, and the length in the in-plane direction satisfies $L_x=H\times N_x/N_y$, where $N_y$ is the number of the cells in $y$ direction.
Set $N_B=20$, and $N_y=100$ or $200$ for the numerical accuracy of the simulation.
In order to capture the non-equilibrium distribution function, $64-1728$ discretized directions are used based on demand.
As $L \leq 10 \text{um}$, let $\beta=0.2$, while as $L=100 {\text{um}}$, set $\beta=5.0$ for the numerical stability.
When $\epsilon_2 < 1.0 \times 10^{-11}$, terminate the iteration.

\begin{figure}
 \centering
 \subfloat[]{\includegraphics[width=0.45\textwidth]{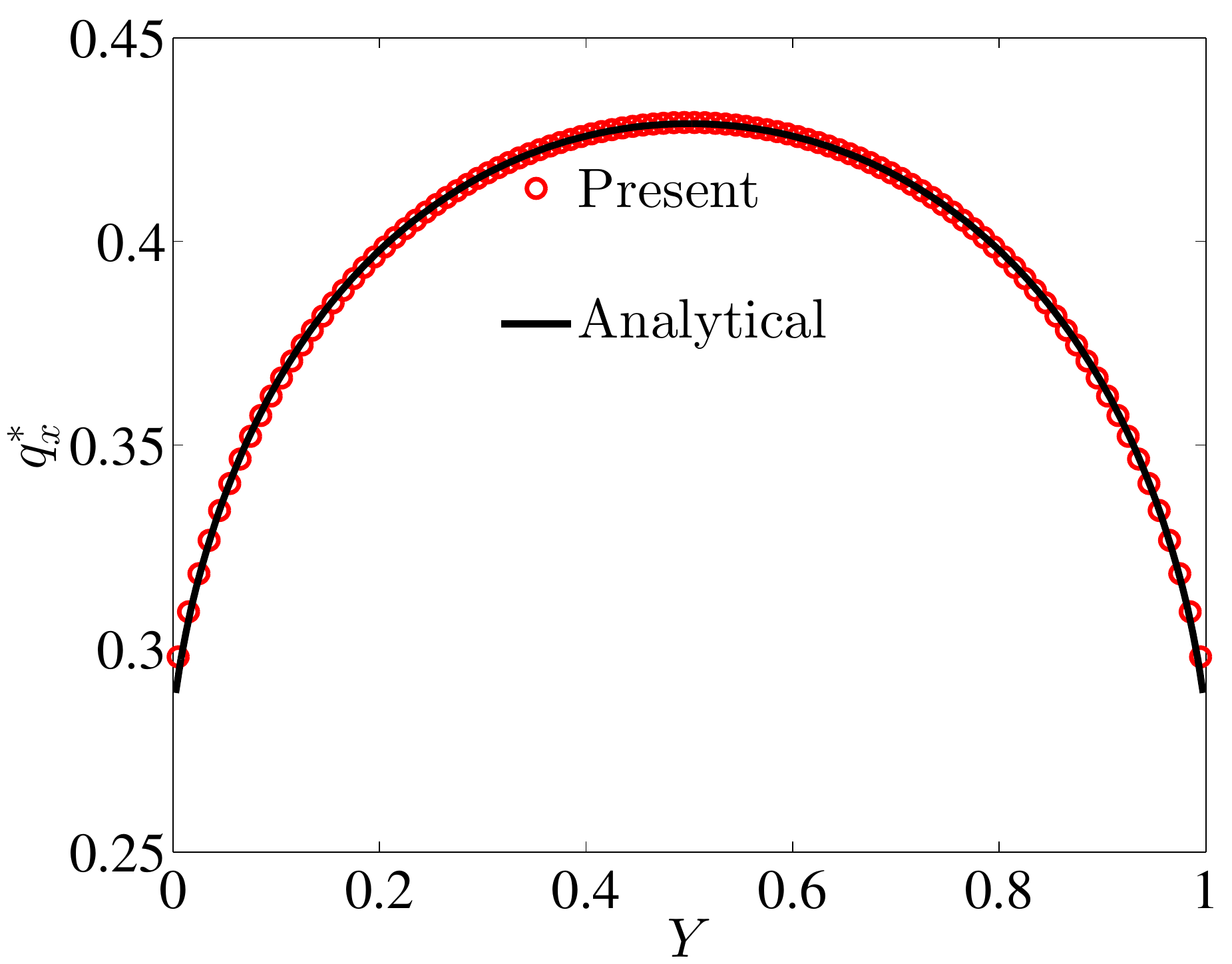}}~~
 \subfloat[]{\includegraphics[width=0.45\textwidth]{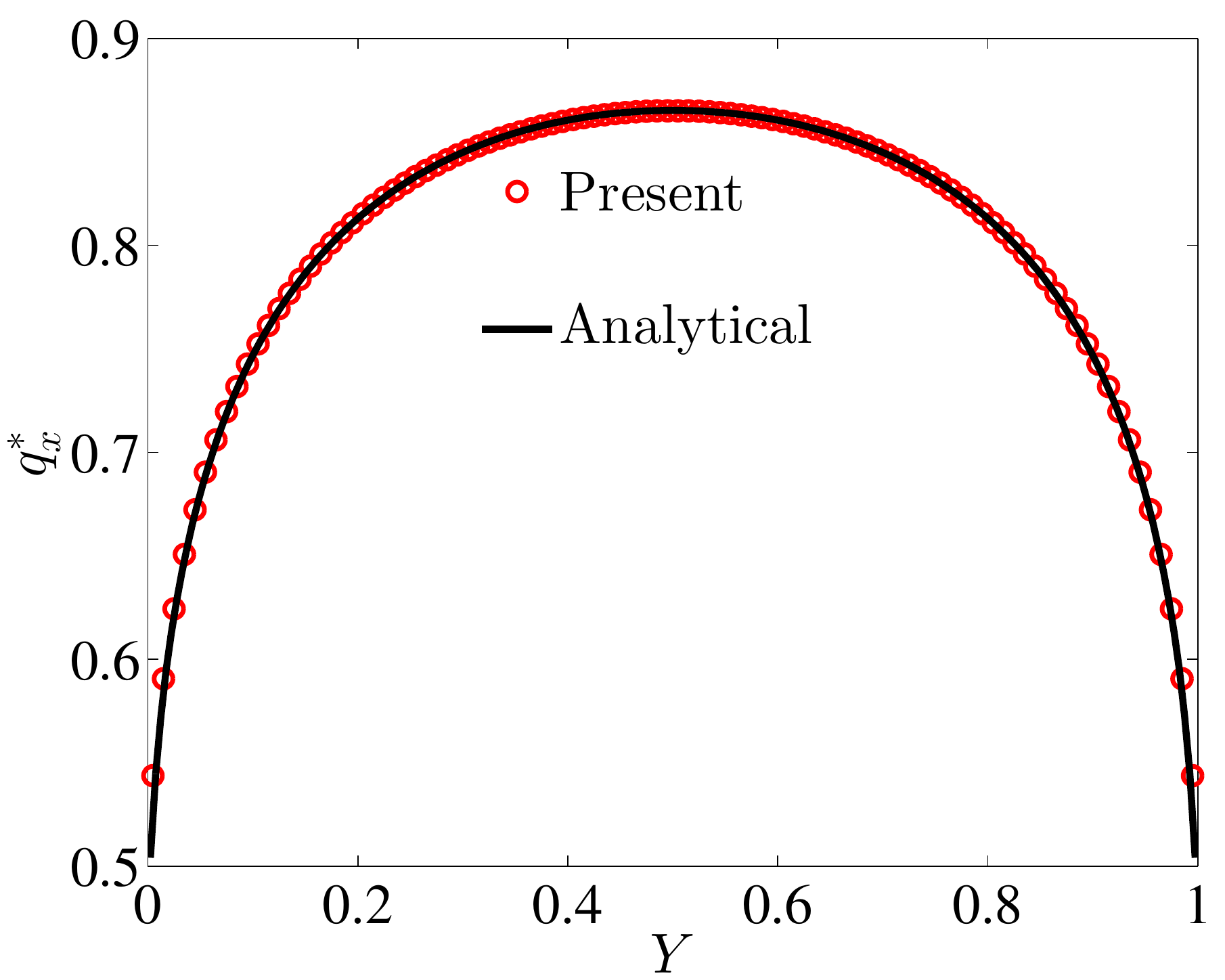}}~~ \\
 \subfloat[]{\includegraphics[width=0.45\textwidth]{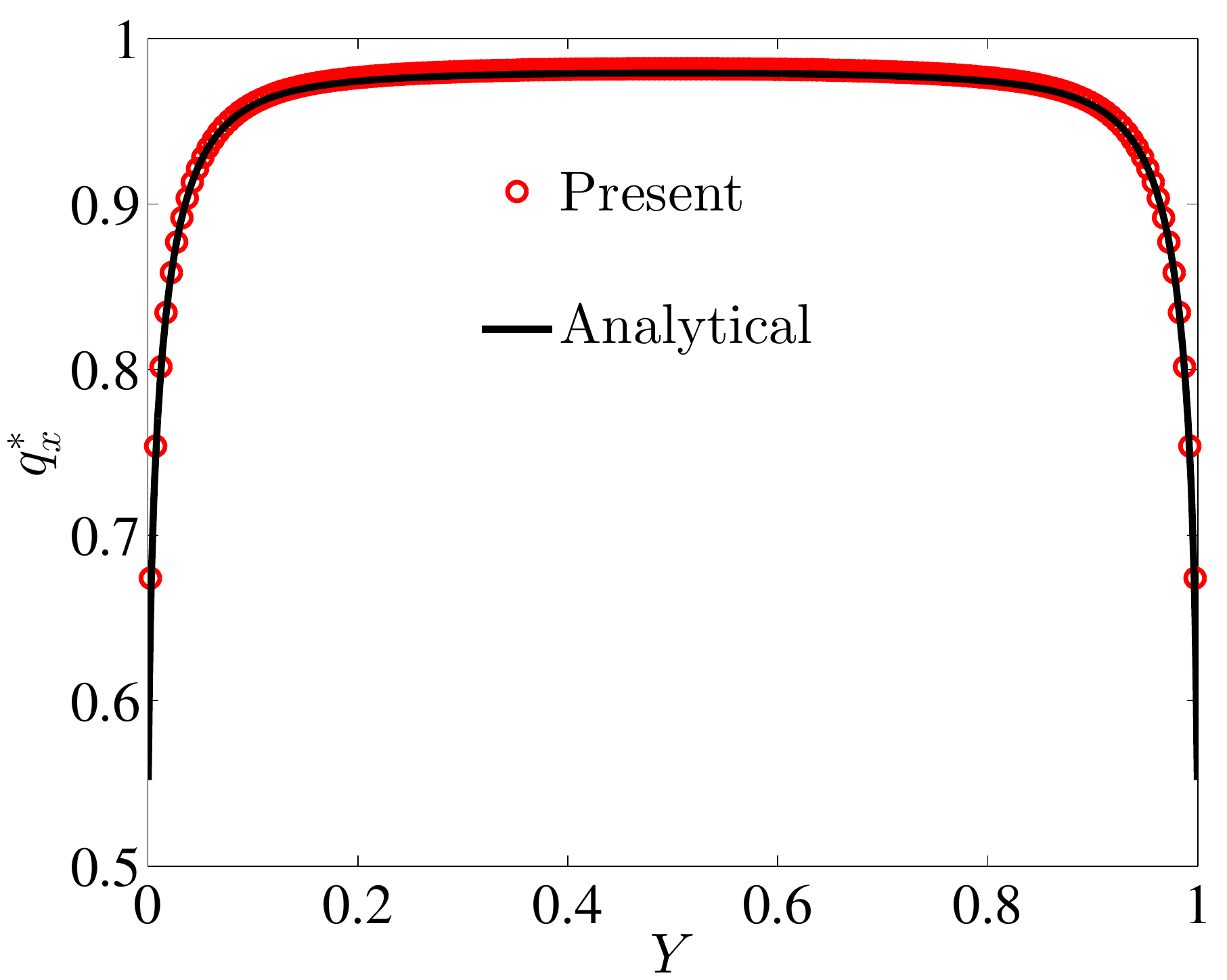}}~~
 \subfloat[]{\includegraphics[width=0.45\textwidth]{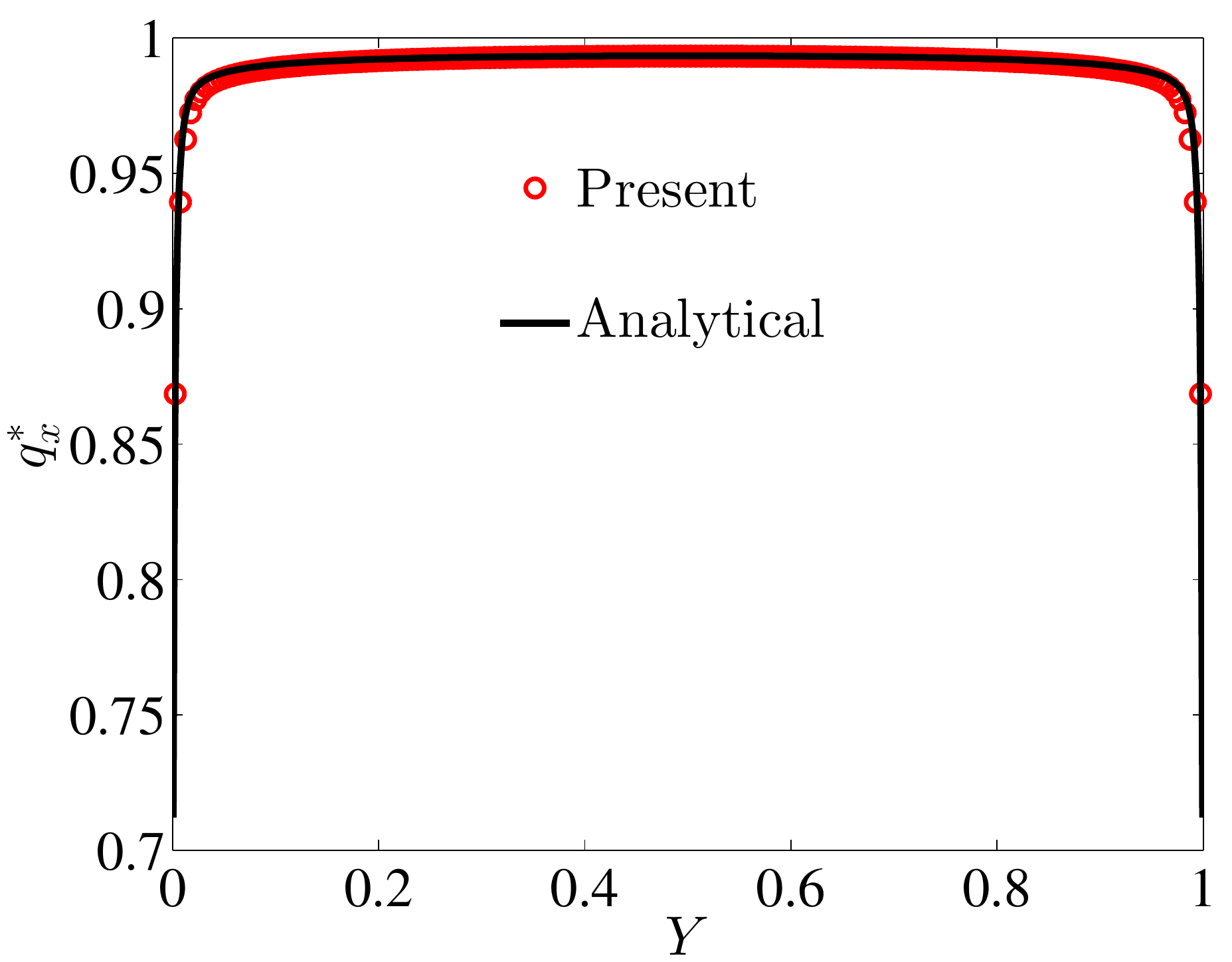}}~~\\
 \caption{The distribution of the non-dimensional $x$-directional heat flux ($q_{x}^{*}=~q_{x}(Y)/q_{bulk}$, where $q_{bulk}=-k_{bulk}\times dT/dx$.) in the $y$ direction ($Y=y/H$) with different distances ($H$) between the two planes. Red circle is the present numerical results and the black solid line is the analytical solution. (a) $H=$ 100nm, $N_y=100$, $N_{\theta}=72,~N_{\varphi}=24$, (b) $H=$ 1um, $N_y=100$, $N_{\theta}=24,~N_{\varphi}=16$ (c) $H=$ 10um, $N_y=200$, $N_{\theta}=24,~N_{\varphi}=16$, (d)$H=$ 100um, $N_y=200$, $N_{\theta}=8,~N_{\varphi}=8$.}
 \label{heatfluxinplane}
\end{figure}

\begin{figure}
 \centering
 \includegraphics[width=0.45\textwidth]{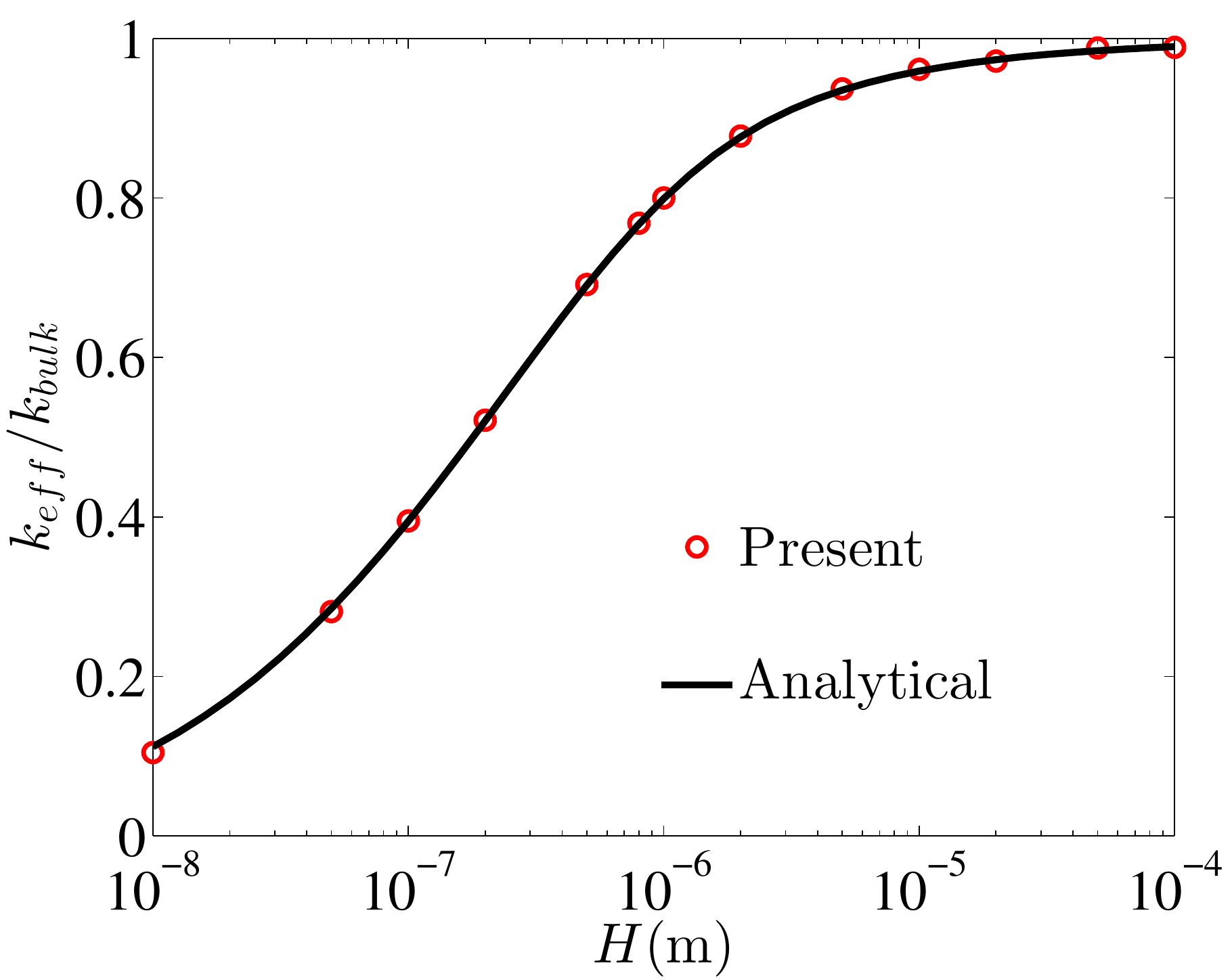}~
 \caption{The effective thermal conductivity ($k_{eff}=-(dT/dx)^{-1} \int_{0}^{1} q_{x}(Y)dY$) with different distances ($H$) between the two planes. Red circle is the present numerical results and the black solid line is the analytical solution.}
 \label{keffinplane}
\end{figure}

Figure~\ref{heatfluxinplane} shows the distribution of the non-dimensional $x$-directional heat flux ($q_{x}^{*}=~q_{x}(Y)/q_{bulk}$, where $q_{bulk}=-k_{bulk}\times dT/dx$.) in the $y$ direction ($Y$) with different distances ($H$) between the two planes.
It can be observed that present numerical results are in good agreement with the analytical solutions over a large length scales except some deviations near the boundary.
The effective thermal conductivity ($k_{eff}=-(dT/dx)^{-1} \int_{0}^{1} q_{x}(Y)dY$) in the whole length range predicted by the present scheme is also compared with the analytical solution, as shown in~\cref{keffinplane}.
Although there are some discrepancies when $H<$ 100 nm due to the highly non-equilibrium effects, generally speaking, the data obtained by the present scheme is consistent with the analytical solution.
Besides, when $H=100$ um, it takes about 3 minutes to obtain converged solution for the present scheme, while the implicit DOM costs about 26 minutes on the same comupter (wall time; message passing interface (MPI) direction-based parallelization with 8 cores).
For each case simulated in this subsection, the total memory that the present scheme requires is less than 10 megabytes (MB), which is almost on the same order as that of the Fourier solver.


\subsection{2D square periodic pores heat transfer}

\begin{figure}
 \centering
 \includegraphics[width=0.8\textwidth]{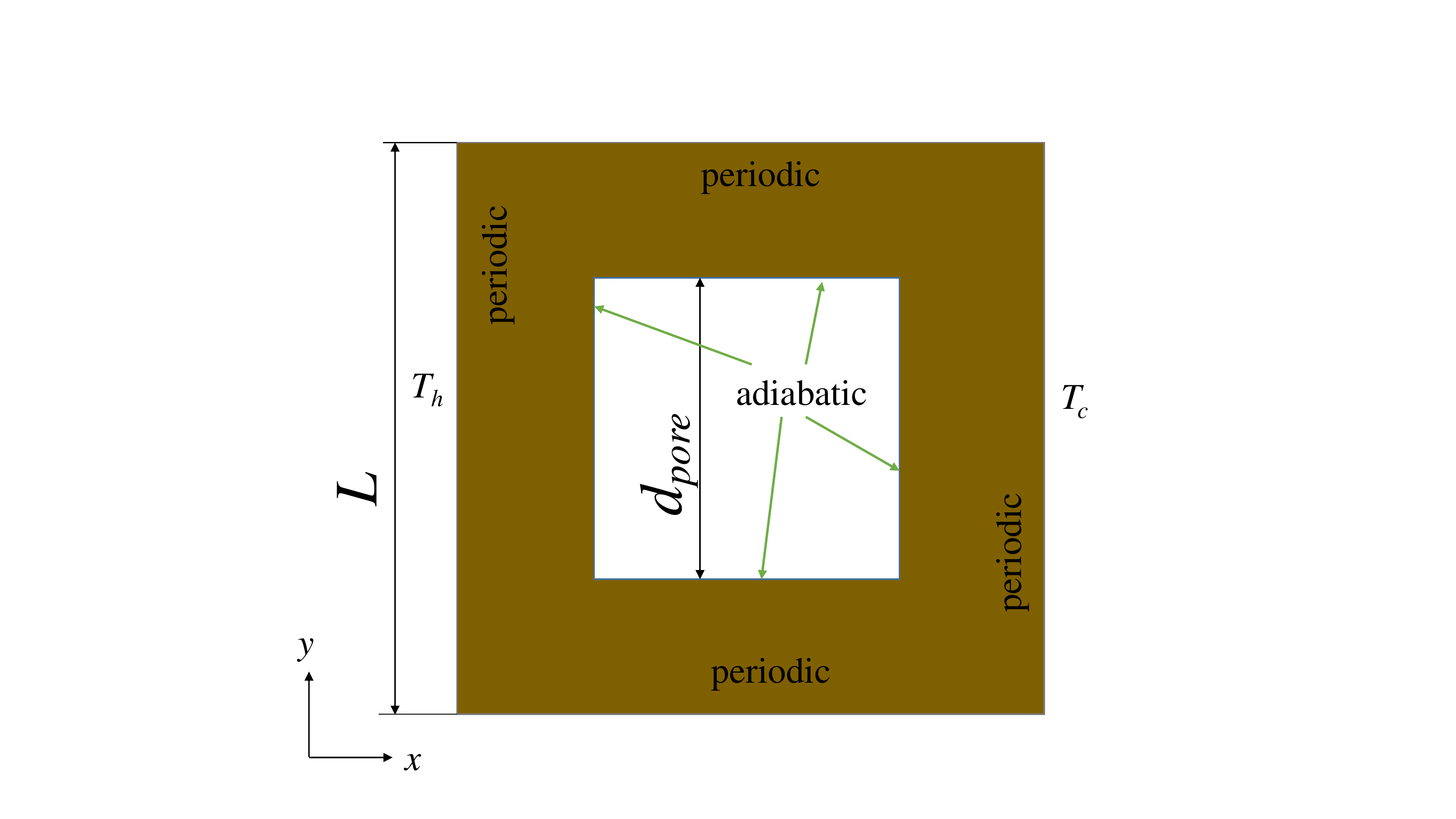}~
 \caption{Schematic of 2D square pore heat transfer.}
 \label{squarepore}
\end{figure}
\begin{figure}
 \centering
 {\includegraphics[width=0.8\textwidth]{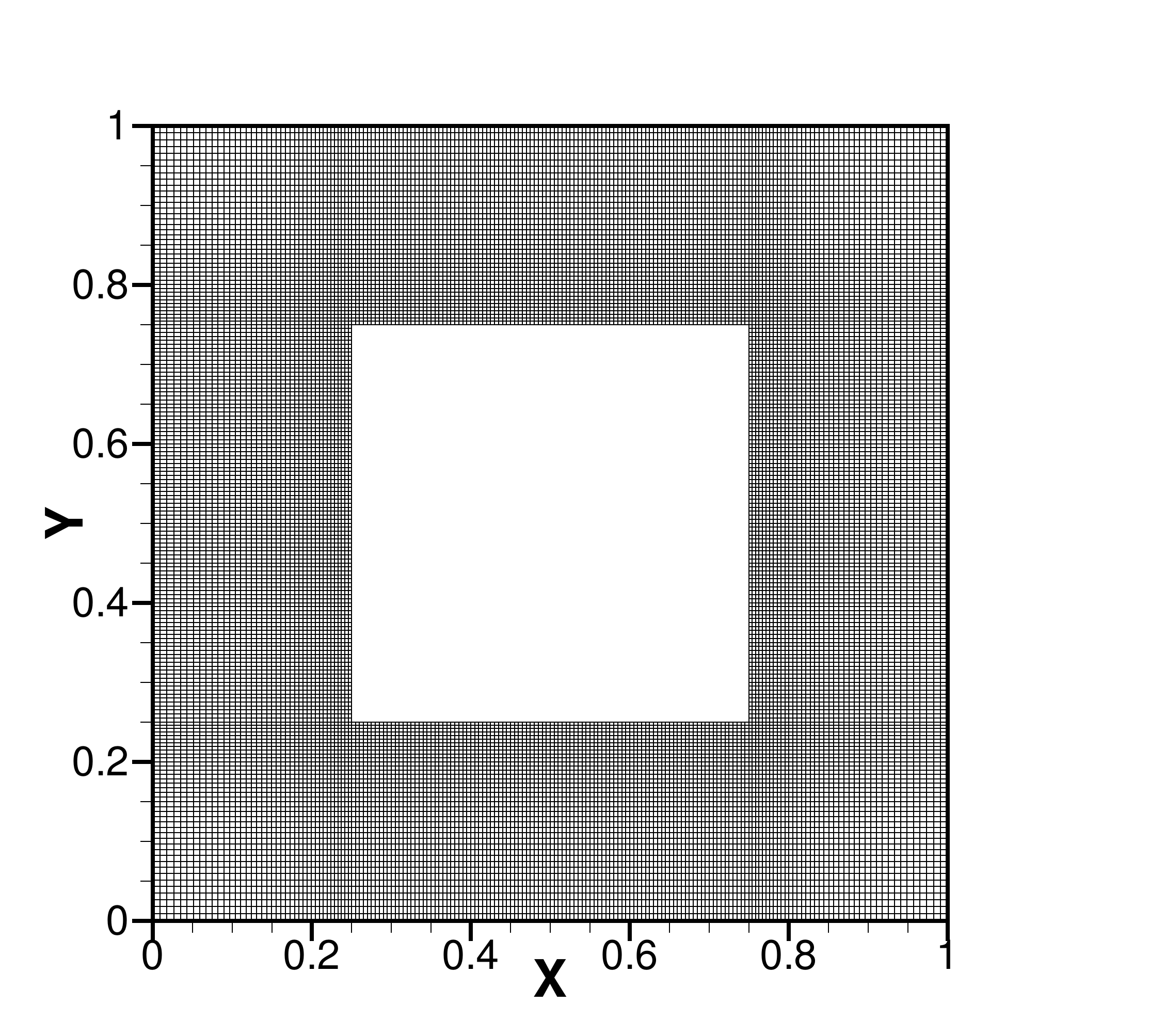}}~~
 \caption{Grid system of the 2D square periodic pores with the porosity $\Phi=0.25$. The normalized length ${\text{X}}=x/L,~{\text{Y}}=y/L$. The number of the total grids is $N_{cell}=180^2-100^2=22400$.}
 \label{meshsystem}
\end{figure}
\begin{figure}
 \centering
 \subfloat[]{\includegraphics[width=0.33\textwidth]{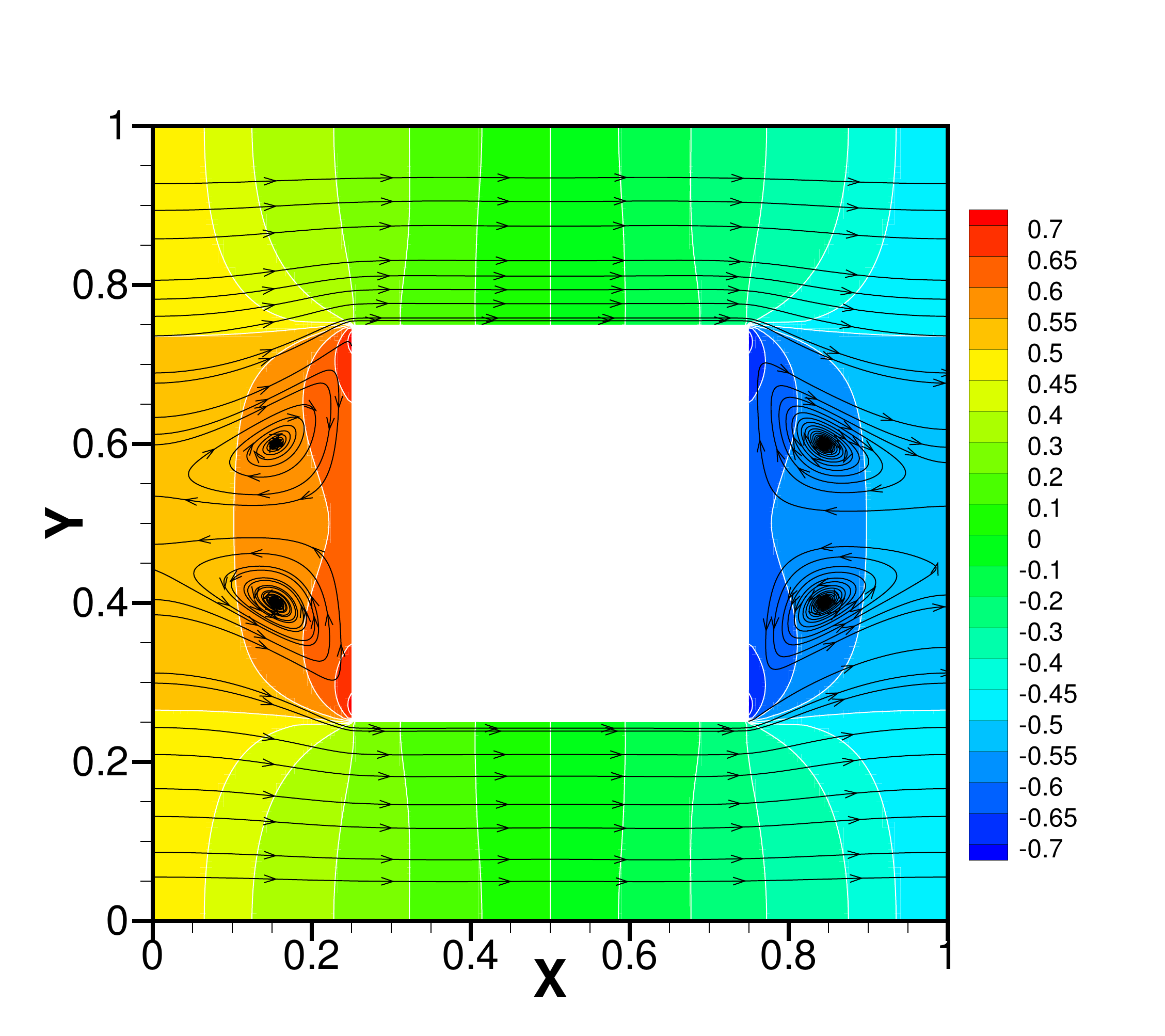}}~~
 \subfloat[]{\includegraphics[width=0.33\textwidth]{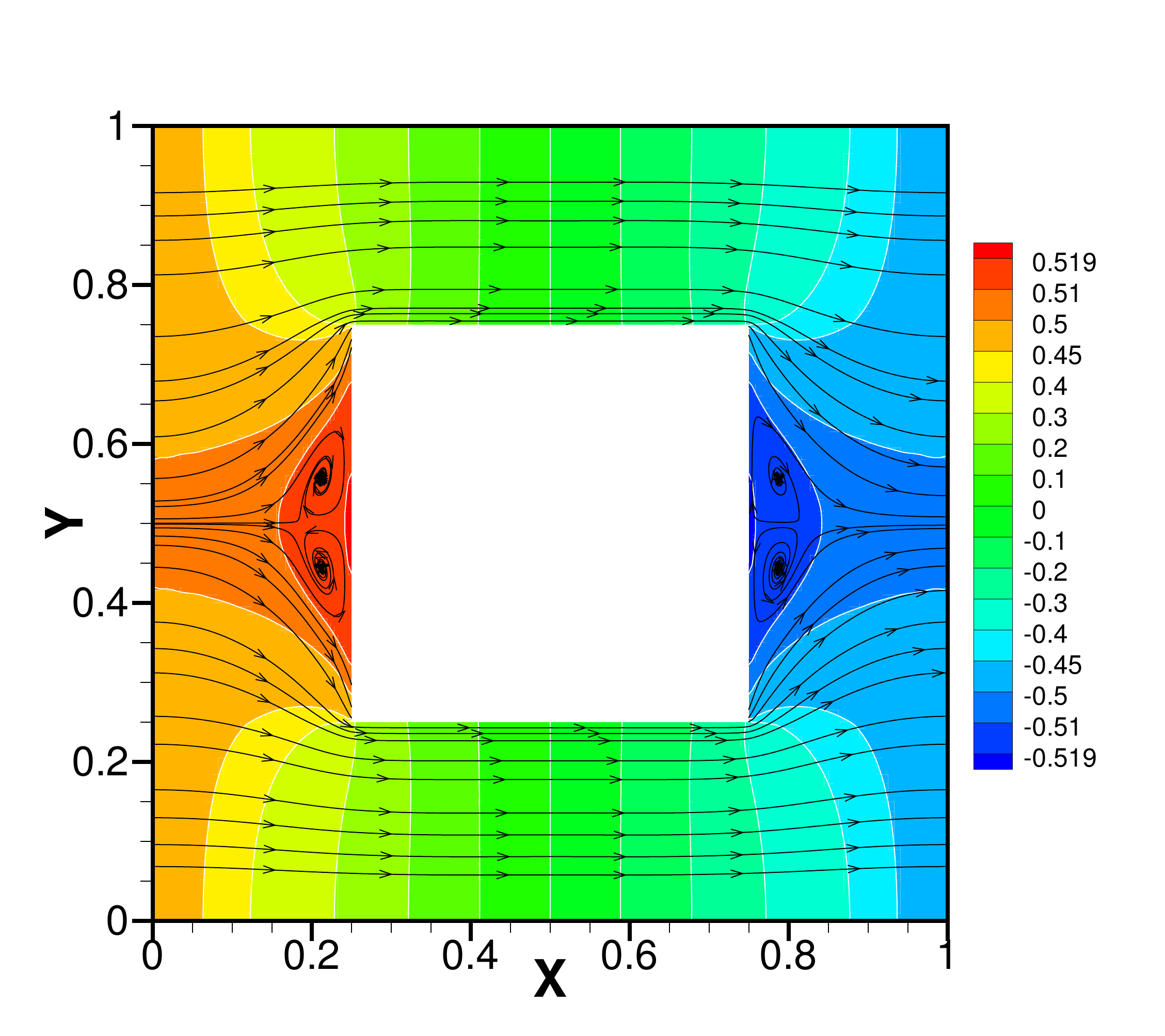}}~~
 \subfloat[]{\includegraphics[width=0.33\textwidth]{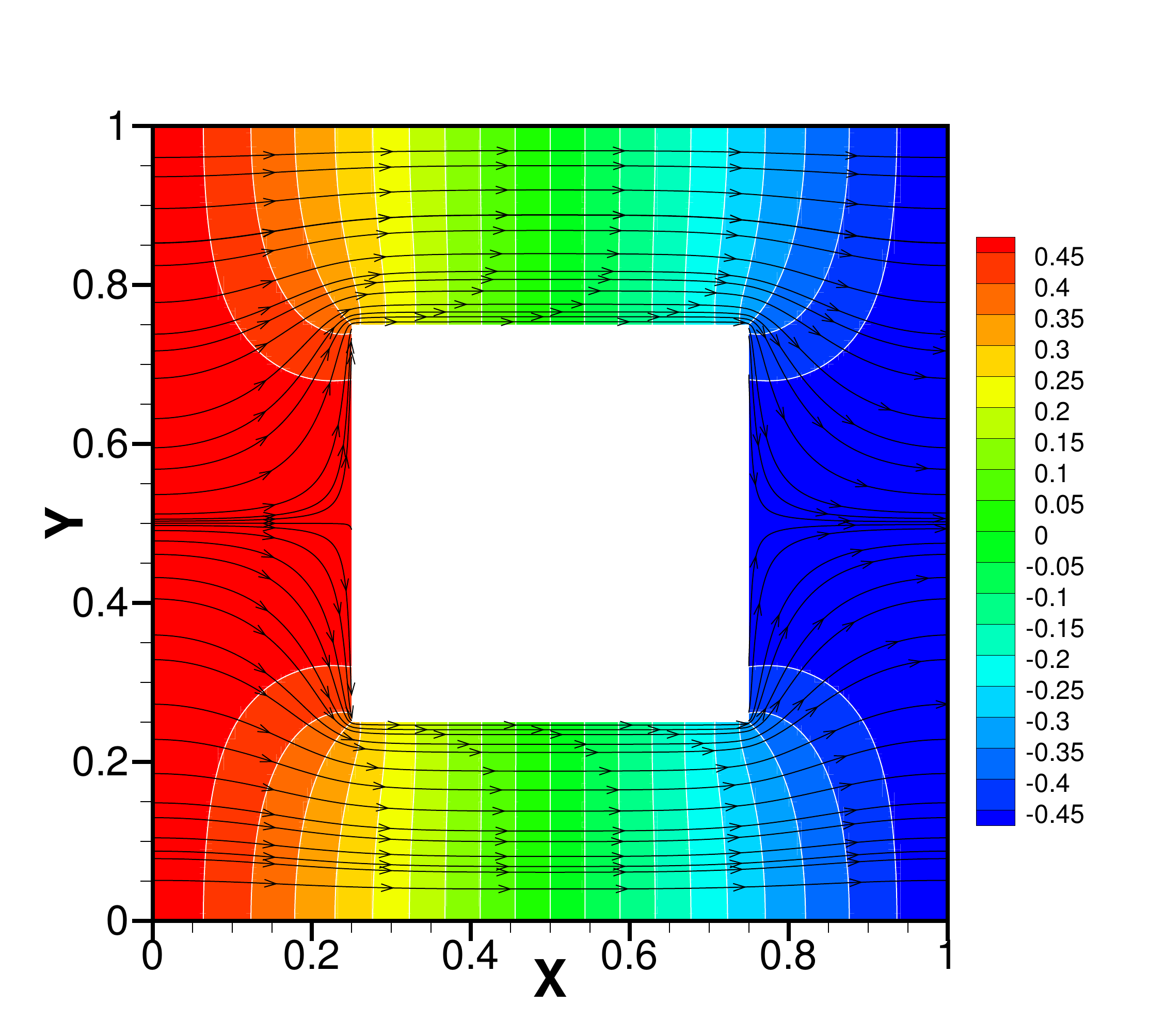}}~~ \\
 \subfloat[]{\includegraphics[width=0.33\textwidth]{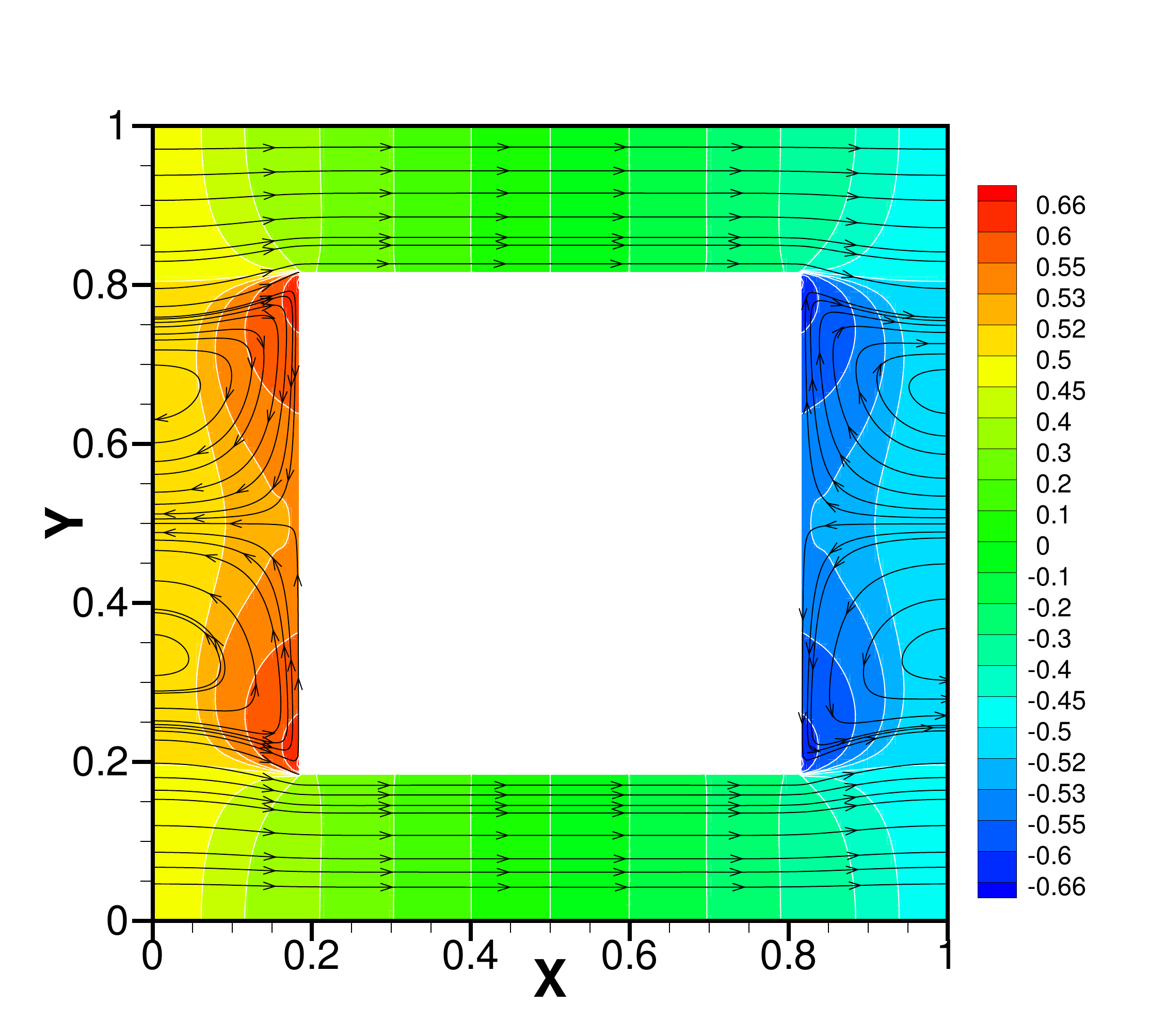}}~~
 \subfloat[]{\includegraphics[width=0.33\textwidth]{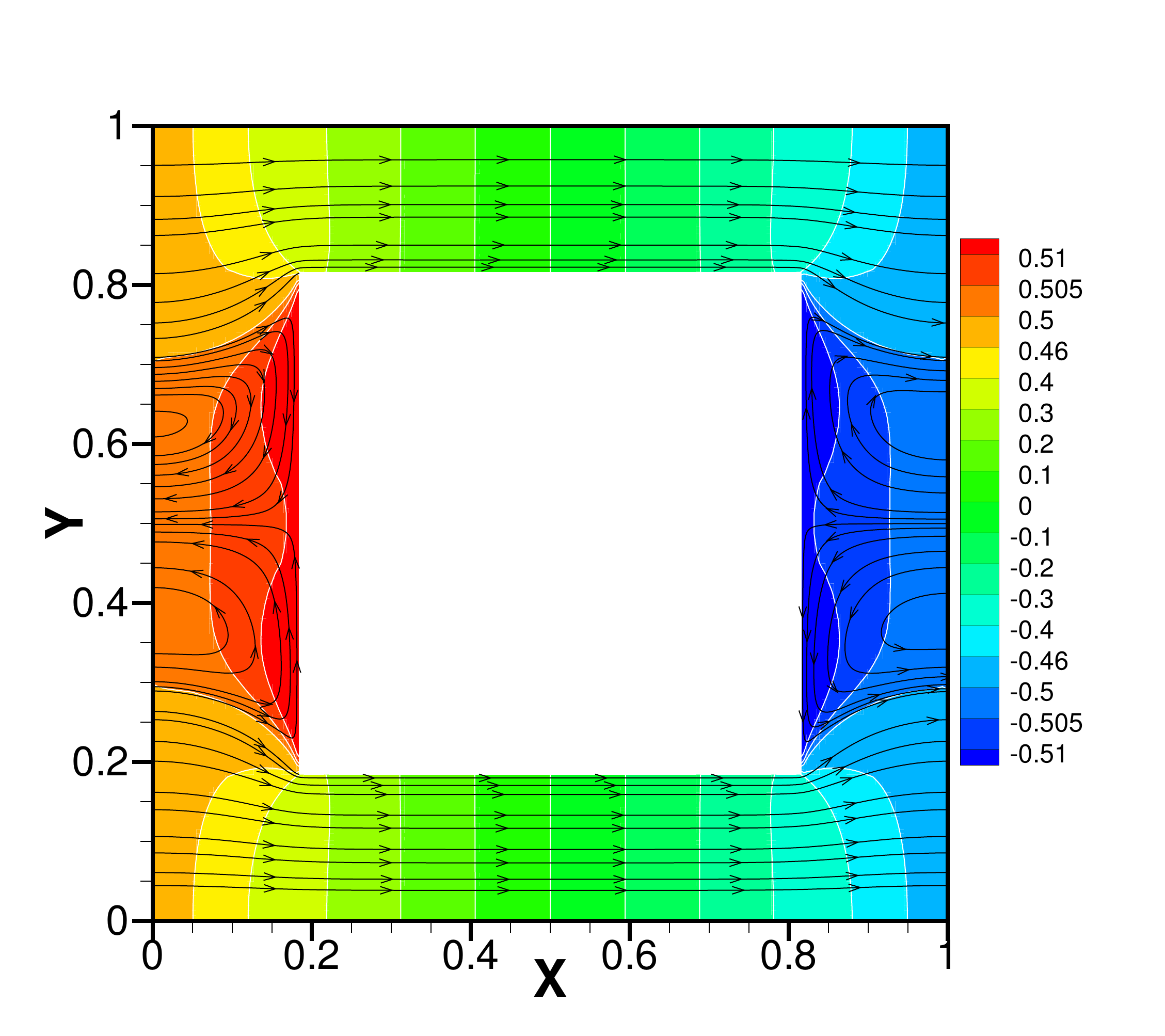}}~~
 \subfloat[]{\includegraphics[width=0.33\textwidth]{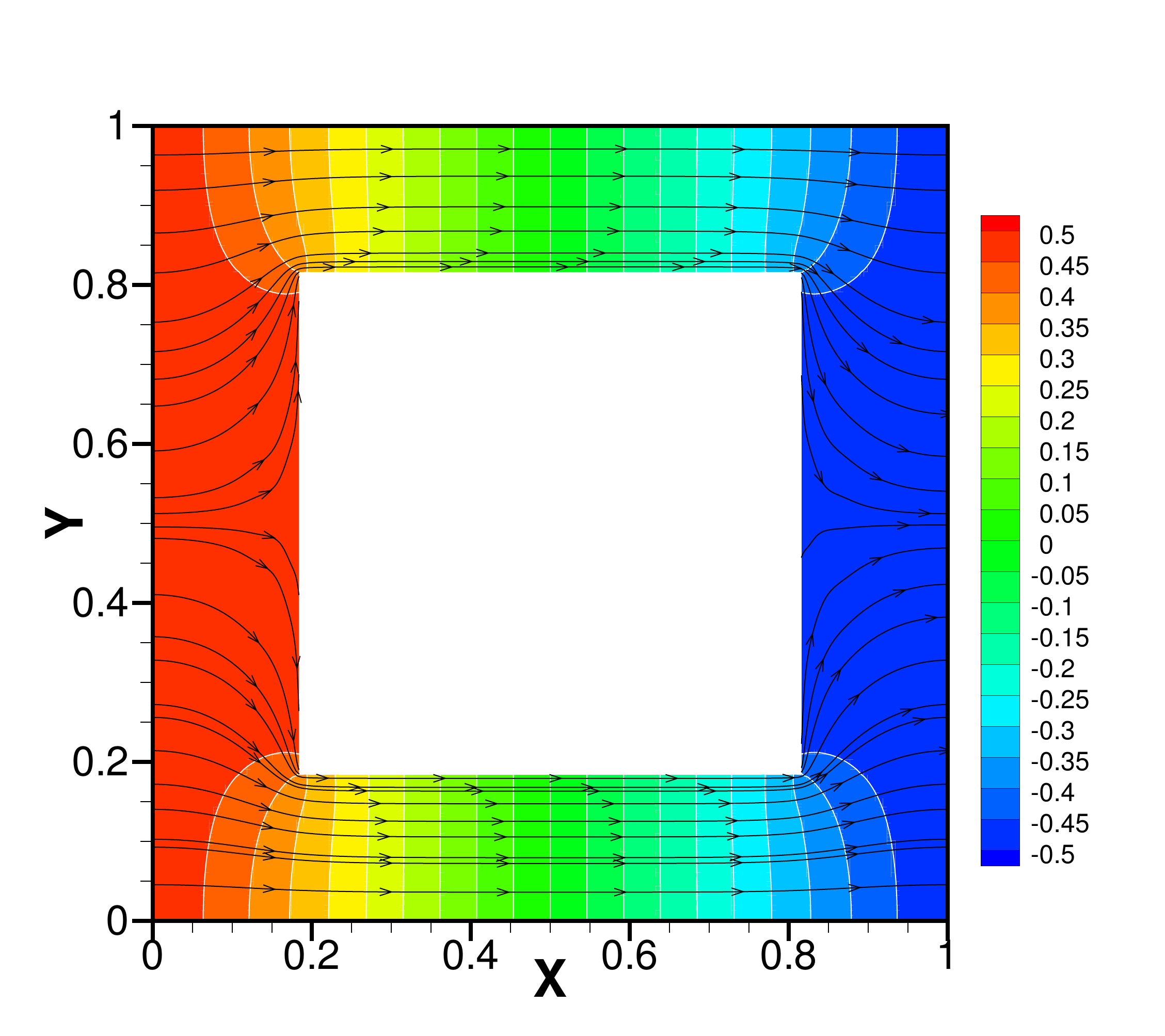}}~~ \\
 \caption{The non-dimensional temperature contour ($T^{*}=(T-T_{\text{ref}})/DT$) and the streamlines of the heat flux of the 2D square periodic pores with the different porosity ($\Phi$) and lengths ($L$), where the normalized length ${\text{X}}=x/L,~{\text{Y}}=y/L$. The colored background and white solid line is the non-dimensional temperature field, and the black line with arrow is the streamlines of the heat flux. (a) $\Phi=0.25$, $L=$10 nm, $N_{\theta}=48,~N_{\varphi}=24$, (b) $\Phi=0.25$, $L=$100 nm, $N_{\theta}=36,~N_{\varphi}=24$, (c) $\Phi=0.25$, $L=$1 um, $N_{\theta}=24,~N_{\varphi}=12$; (d) $\Phi=0.4$, $L=$10 nm, $N_{\theta}=48,~N_{\varphi}=24$, (e) $\Phi=0.4$, $L=$100 nm, $N_{\theta}=36,~N_{\varphi}=24$, (f) $\Phi=0.4$, $L=$1 um, $N_{\theta}=24,~N_{\varphi}=12$.}
 \label{Tporosity}
\end{figure}
\begin{figure}
 \centering
 \subfloat[]{\includegraphics[width=0.45\textwidth]{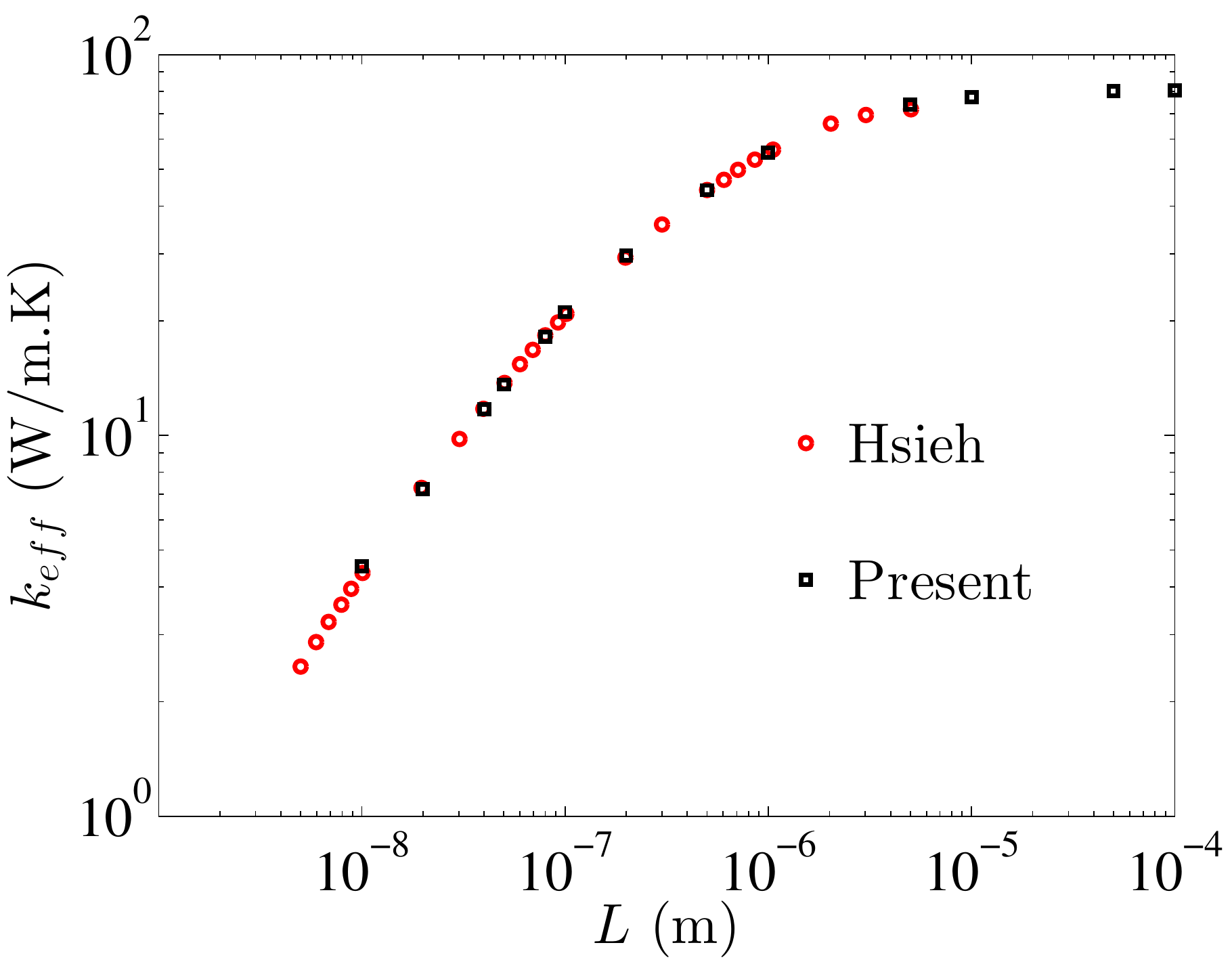}}~~
 \subfloat[]{\includegraphics[width=0.45\textwidth]{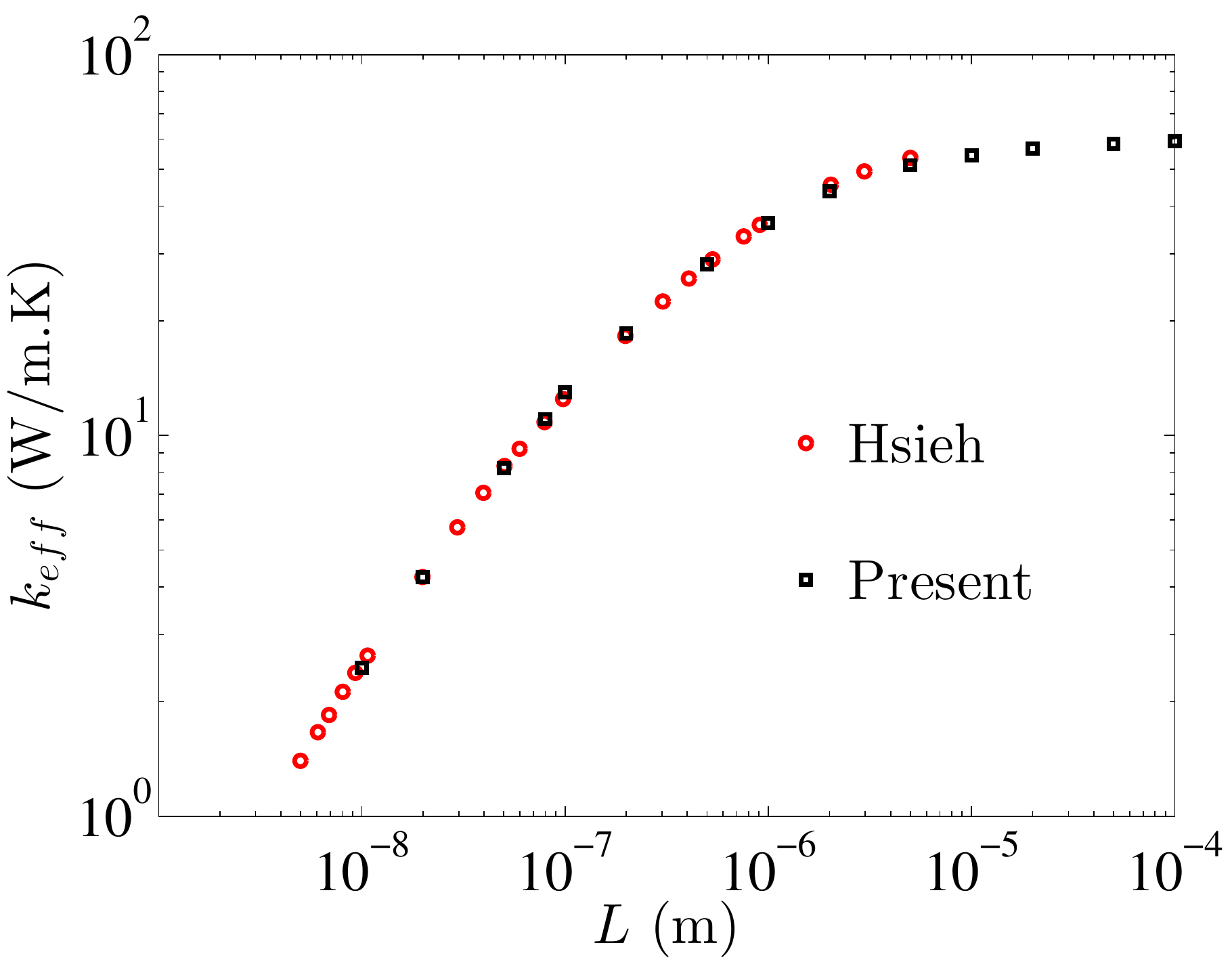}}~~\\
 \caption{The effective thermal conductivity ($k_{eff}$) of the 2D square periodic silicon pore with different lengths ($L$) and porosities ($\Phi$). The red circles are the data obtained from the Hsieh's paper~\cite{hsieh2012thermal}, the black squares are the present numerical results. (a) $\Phi=0.25$, (b) $\Phi=0.40$.}
 \label{keffporosity}
\end{figure}
\begin{figure}
 \centering
 \subfloat[]{\includegraphics[width=0.45\textwidth]{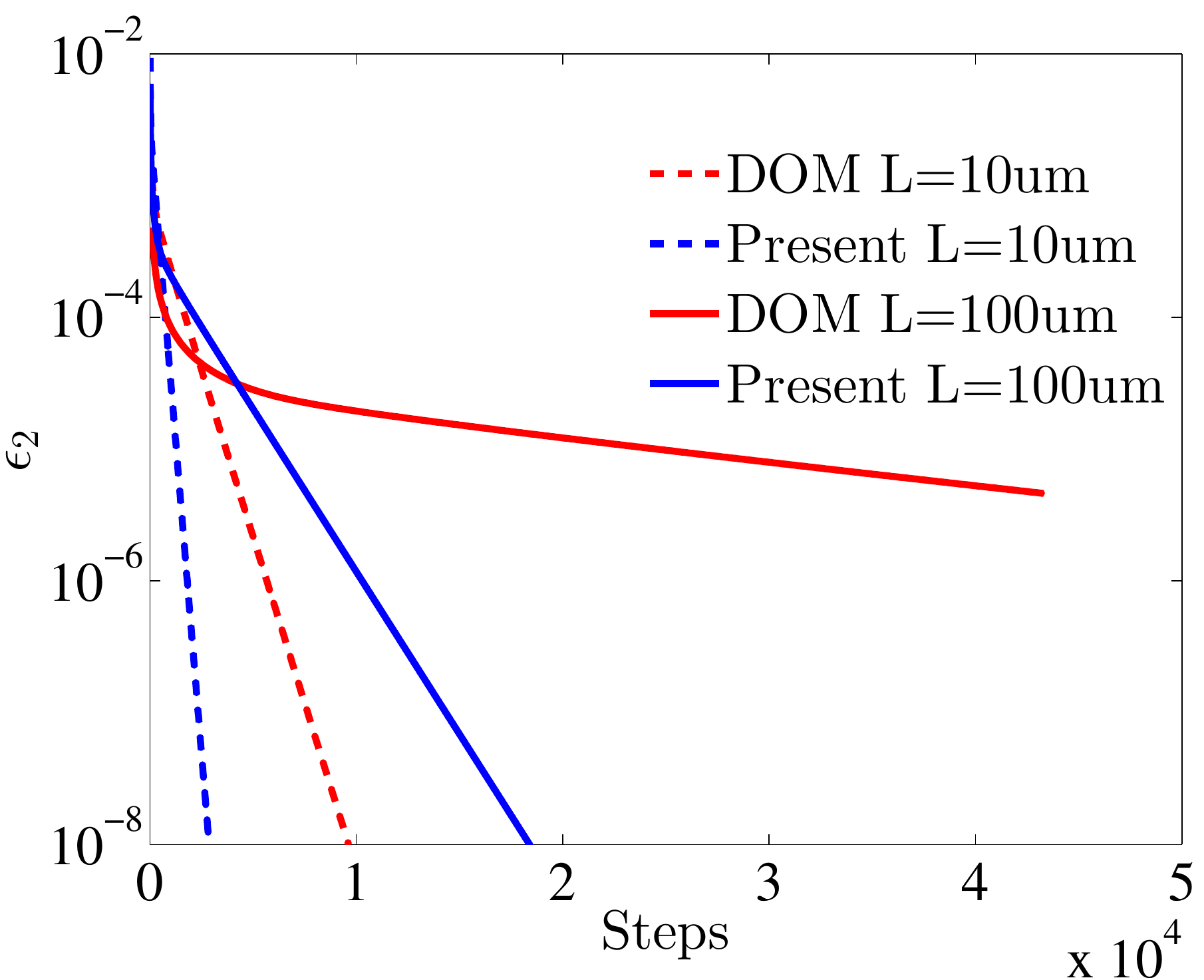}}~~
 \subfloat[]{\includegraphics[width=0.45\textwidth]{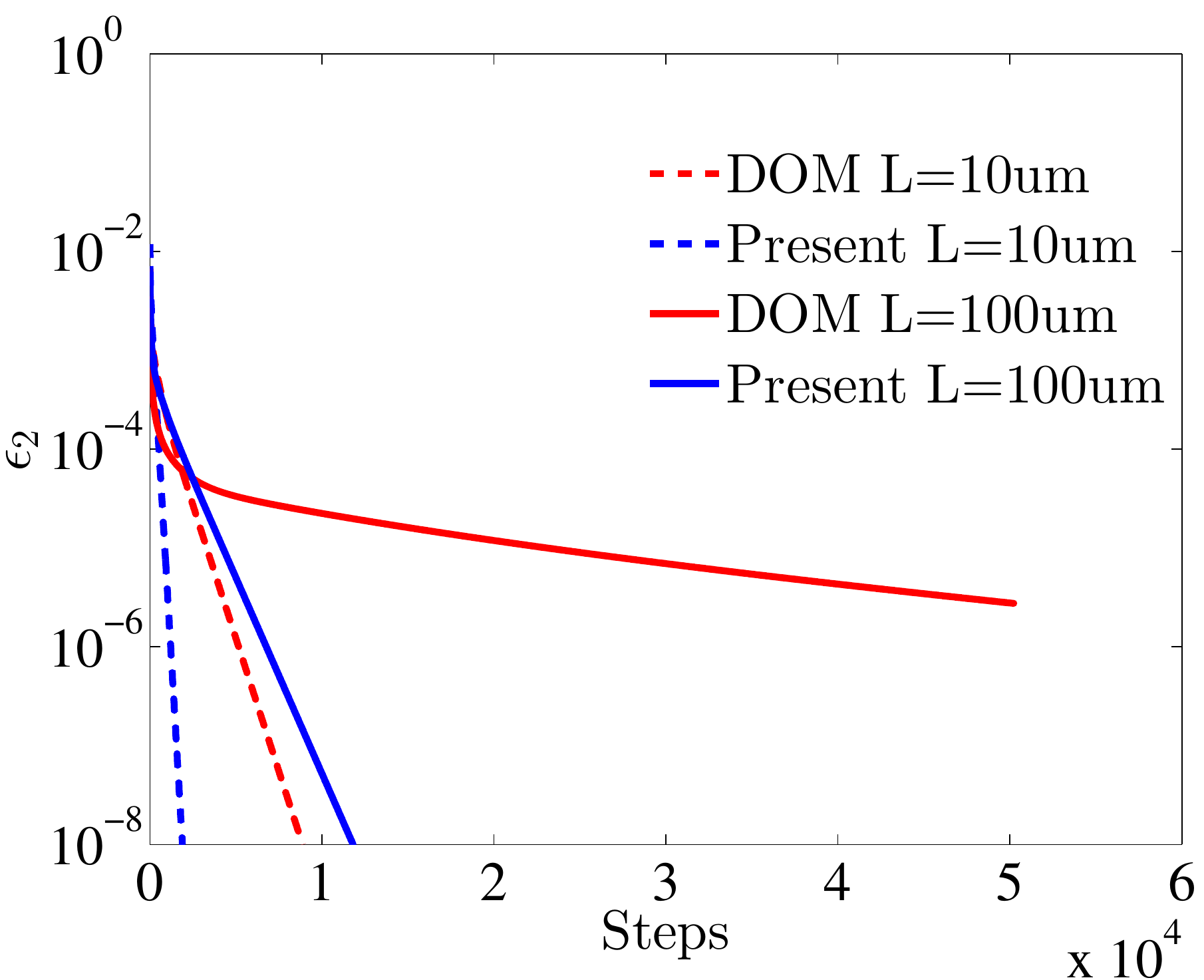}}~~\\
 \caption{The history of the residual $\epsilon_2$ between the present scheme and the implicit DOM with different lengths ($L=$10 um, 100um) and porosities ($\Phi=0.25,~0.4$). Red is the implicit DOM, while the blue is the present scheme; dashed line is $L=10$ um, while the solid line is $L=$100 um. (a) $\Phi=0.25$ ($L=10$ um, $\beta=1.0$; $L=100$ um, $\beta=5.0$), (b) $\Phi=0.40$ ($L=10$ um, $\beta=0.20$; $L=100$ um, $\beta=4.0$).}
 \label{epsilonporosity}
\end{figure}

In recent years, the nanoporous materials get a lot of attention for its application in thermoelectric conversion~\cite{Bera10nanoporous,hsieh2012thermal,LIANG20174915}.
In this work, the transverse heat transfer in the 2D square periodic pores (\cref{squarepore}) is investigated numerically.
$d_{pore}$ is the side length of the square pore and the $L$ is the length of the periodic square unit cell (the pore and the unit cell are concentric, and $L_{\text{ref}}=L$).
Define the porosity ($\Phi$) of the unit cell as $\Phi=S_{pore}/S_{unit}=(d_{pore}/L)^2$.
A constant temperature difference ($T_h-T_c=DT$, where $T_h=T_{\text{ref}}+DT/2,~T_c=T_{\text{ref}}-DT/2$) is added across the $x$ directional boundary of the unit cell (heat flows from left to right), while there is no temperature difference across the $y$ directional boundary.
The periodic boundary condition is implemented on the left and right boundary of the unit cell, as well as the top and bottom boundary.
The diffusely reflecting boundary condition is adopted for the whole boundary of the square pore.
Set $N_B=20$, and $64-1152$ discretized directions are used to accurately capture the non-equilibrium effects in different scales based on demand.
Let $\beta=5.0$ when $L=100 {\text{um}}$, and when $L \leq 10{\text{um}}$, set $\beta=0.2$.
The heat transfer phenomena with different porosities ($\Phi=0.25,~0.40$) and lengths ($L$) of the unit cell are studied.
Figure~\ref{meshsystem} shows the non-uniform grid system of the 2D square periodic pores with the porosity $\Phi=0.25$, the total number of the cells is $N_{cell}=180^2-100^2=22400$.
Similar grid system is implemented when $\Phi=0.40$.
When $\epsilon_2 < 1.0 \times 10^{-8}$, stop the whole iteration.

Figure~\ref{Tporosity} shows the non-dimensional temperature contour ($T^{*}=(T-T_{\text{ref}})/DT$) and the streamlines of the heat flux of the 2D square periodic pores with the different porosities ($\Phi=0.25,~0.40$) and lengths ($L=$ 10nm, 100nm, 1um).
When $L \leq$ 100 nm, it can be observed that the temperature is relatively higher on the left area of the square pore while lower on the right area.
It's because that, as the characteristic length decreases, the effects of the boundary scattering increase and play a main role of the phonon transport.
On the left area of the square pore, the phonons transport from the high temperature areas to the pore boundary and the phonons reflected from the boundary stack together, which results in that more and more phonons arrive.
On the contrary, few phonons can arrive on the right of the square pore.
Besides, it can be observed that the heat flows from low temperature to high temperature in some highly non-equilibrium areas, which indicates the failure of the Fourier law at the micro/nano scale.
When $L \geq $ 1 um, the phonon intrinsic scattering becomes dominated.
As a result, the heat flow phenomena come closer to that in the diffusive regime.
The effective thermal conductivity ($k_{eff}$) at different length scales predicted by the present scheme keeps consistent with the Hsieh's results~\cite{hsieh2012thermal}, as shown in~\cref{keffporosity}, where
\begin{equation}
k_{eff}=  \overline{q_{x}} \frac{L}{DT},
\end{equation}
where $\overline{q_{x}}$ is the average value of the $x$ directional heat flux over the whole unit cell, i.e.,
\begin{equation}
\overline{q_{x}}=\frac{1}{L^2} \int_{0}^{L} \int_{0}^{L} q_{x}(x,y)dxdy.
\end{equation}
Figure~\ref{epsilonporosity} shows the history of the residual $\epsilon_2$  between the present scheme and the implicit DOM with different lengths ($L=$10 um, 100um) and porosities ($\Phi=0.25,~0.4$).
It can be observed that through adding the macroscopic iteration, the present scheme accelerates convergence significantly compared to the implicit DOM.

\section{Conclusion}

In the present study, an efficient implicit kinetic scheme based on the non-gray model including the phonon dispersion and polarization was developed for multiscale heat transfer problem.
In order to realize the efficient coupling of phonons with different mean free paths, the present scheme takes account of the phonon transport behaviors at the microscopic and macroscopic levels simultaneously.
At the microscopic level, the microscopic iteration is used to capture the granularity of phonon behavior through solving the phonon BTEs under the discretized wave vector space in turn.
At the macroscopic level, a macroscopic equation is introduced and solved iteratively to represent the collective effect of the phonon.
The bridge connecting the microscopic iteration and the macroscopic iteration is the total heat flux, which is obtained from the moments of the phonon distribution function over all frequencies, branches and directions.
The numerical iteration and physical information exchange at two different scales realize the efficient phonon coupling both in the spatial space and the wave vector space and make the present scheme very fast and economic for multiscale heat transfer problem in a wide range.
%

Three numerical tests were presented and validate that the present scheme can capture the multiscale heat transfer phenomena accurately in a wide length and temperature range.
With the consideration of the phonon dispersion and polarization, the predictions of the present scheme agree with the experimental data in the cross-plane or the analytical solutions in the in-plane heat transfer.
Compared to the implicit DOM, the present scheme with the macroscopic iteration accelerates convergence by tens of times.
In addition, even if including the phonon dispersion and polarization, the memory cost of the present scheme is still on the same order as the Fourier solver.
In consideration of the superior performance, the present scheme is an efficient tool to simulate the multiscale heat transfer in micro/nanomaterials.
%

\section*{Acknowledgments}

This work was supported by the National Key Research and Development Plan (No. 2016YFB0600805) and the National Science Foundation of China (11602091, 91530319).

\section*{References}
\bibliographystyle{IEEEtr}
\bibliography{zhangcnongray}

\end{document}